\documentclass[journal]{IEEEtranTCOM}

\normalsize
\ifCLASSINFOpdf
\else
\fi

\usepackage{amsmath,amsfonts,amssymb,varioref,mathrsfs,verbatim}
\usepackage{graphics}
\usepackage{graphicx}
\usepackage{epsfig}
\usepackage{subcaption}

\usepackage[font=small,labelfont=bf, figurename=Fig.]{caption} 
\usepackage{color}
\usepackage[normalem]{ulem}

\begin{document}
%
\title{Compressive Sensing for Millimeter Wave  \\ Antenna Array Diagnosis}
%
%
%

\author{Mohammed E. Eltayeb, Tareq Y. Al-Naffouri, and Robert W. Heath, Jr. 
\thanks{Part of this work was presented in the 2016 IEEE Global Communications Conference  \cite{mg1}. Mohammed E. Eltayeb  is with the California State University, Sacramento, California, USA (e-mail: mohammed.eltayeb@csus.edu),  Robert W. Heath, Jr. is with the University of Texas at Austin, USA (email: rheath@utexas.edu), and
Tareq Y. Al-Naffouri is with King Abdullah University of Science and Technology, Saudi Arabia (e-mail:
tareq.alnaffouri@kaust.edu.sa). This research was partially supported
by the U.S. Department of Transportation through the Data-Supported Transportation
Operations and Planning (D-STOP) Tier 1 University Transportation
Center and by the Texas Department of Transportation under Project 0-6877
entitled Communications and Radar-Supported Transportation Operations and
Planning (CAR-STOP).}}


\maketitle

\date{}


\begin{abstract}

The radiation pattern of an antenna array depends on the excitation weights and the geometry of the array. Due to wind and atmospheric conditions, outdoor millimeter wave antenna elements are subject to full or partial blockages from a plethora of particles like dirt, salt, ice, and water droplets.  Handheld devices are also subject to blockages from random finger placement and/or finger prints. These blockages cause absorption and scattering to the signal incident on the array,  modify the array geometry, and  distort  the far-field radiation pattern of the array. This paper studies the effects of blockages on the far-field radiation pattern of linear arrays and proposes several array diagnosis techniques for millimeter wave  antenna arrays. The proposed techniques jointly estimate the locations of the blocked antennas and the induced attenuation and phase-shifts given knowledge of the angles of arrival/departure.  Numerical results show that the proposed techniques provide satisfactory results in terms of fault detection with reduced number of measurements (diagnosis time) provided that the number of blockages is small compared to the array size.
\end{abstract}


\IEEEpeerreviewmaketitle

\section{Introduction} \label{sec:Intro}
\IEEEPARstart{T}{he} abundance of bandwidth in the millimeter wave (mmWave) spectrum enables gigabit-per-second data rates for cellular systems and local area networks \cite{m1}, \cite{m2}. MmWave systems make use of large antenna arrays at both the transmitter and the receiver  to provide sufficient receive signal power. The use of large antenna arrays is justified by the small carrier wavelength at mmWave frequencies which permits large number of antennas to be packed in small form factors.

\begin{figure}[t]
		\begin{center}
\includegraphics[width=2.2in]{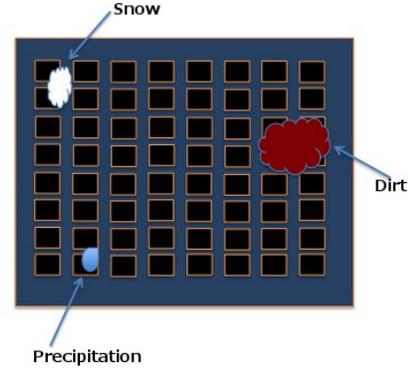}
				\caption{An example of an  outdoor millimeter wave antenna array with different suspended particles partially blocking the array. The suspended particles, with different absorption and scattering properties, modify the array geometry.}
				\label{fig:ant}
		\end{center}
	\end{figure}

Due to weather and atmospheric effects, outdoor mmWave antenna elements are subject to blockages from flying debris or particles found in the air as shown in Fig. \ref{fig:ant}.  The term ``blockage'' here refers to a physical object partially or completely blocking a subset of antenna elements and should not be confused with mmWave channel blockage. MmWave antennas on handheld devices are also subject to blockage from random finger placement and/or fingerprints on the antenna array. Partial or complete blockage of some of the antenna elements reduces the amount of energy incident on the antenna \cite{absorb}, \cite{absorb2}.  For instance, it is reported in \cite{absorb3} that  $90\%$ of a 76.5 GHz signal energy will be absorbed by a water droplet of thickness 0.23 mm.  A thin water film caused by, for example a finger print, is also reported to cause attenuation and a phase shift on mmWave signals \cite{absorb3}. Moreover, snowflakes, ice stones, and dry and damp sand particles are  reported to cause attenuation and/or scattering \cite{absorb}-\cite{absorb4}. Because the size of these  suspended  particles is comparable to the signal wavelength and antenna size, random blockages caused by these particles will change the antenna geometry and result in a distorted radiation pattern \cite{ag0}, \cite{ag}. Random changes in the array's radiation pattern causes uncertainties in the mmWave channel.  It is therefore important to continuously monitor the mmWave system,  reveal any abnormalities, and take corrective measures  to maintain efficient operation of the system. This necessitates the design of reliable and low latency array diagnosis techniques that are capable of detecting the blocked antennas and the corresponding signal power loss and/or phase shifts caused by the blocking particles. Once a fault has been detected,  pattern correction techniques proposed in, for example, \cite{ag0}-\cite{ag5} can be employed to calculate new excitation weights for the array.

Several array diagnostic techniques, which are based on genetic algorithms \cite{gen1}, \cite{gen3}, matrix inversion \cite{matrix}, exhaustive search \cite{esearch}, and MUSIC \cite{music}, have been proposed in the literature to identify the locations of faulty antenna elements. These techniques compare the radiation pattern of the array under test (AUT) with the radiation pattern of an ``error free" reference array. For large antenna arrays, the techniques in \cite{gen1}-\cite{music}  require a large number of samples (measurements) to obtain reliable results. To reduce the number of measurements,  compressed sensing (CS) based techniques have recently been proposed in \cite{cs1}-\cite{cs5}. Despite their good performance,  the techniques in \cite{gen1}-\cite{cs5} are primarily designed to detect the sparsity pattern of a failed array, i.e. the locations of the failed antennas and not necessarily the complex blockage coefficients.  Moreover,  the CS diagnosis techniques proposed in \cite{cs1} and \cite{cs2} have the following limitations: (i) They require measurements to be made at multiple receive locations and are not suitable when both the transmitter and the receiver are fixed. (ii) They assume fault-free receive antennas, i.e. faults at the AUT only, however, faults can occur at both the transmitter and the receiver. (iii) They can not exploit correlation between faulty antennas to further reduce the diagnosis time. (iv) They do not estimate the effective antenna element gain, i.e. the induced attenuation and phase shifts caused by blockages. These estimates can be used to re-calibrate the array. v) They do not optimize the CS  measurement matrices, i.e. the restricted  isometry property might not be satisfied. While the CS technique proposed in \cite{cs5} performs joint fault detection/estimation and angle-of-arrival/departure (AoA/D)  estimation, it is not suitable for mmWave systems as it requires a separate RF chain for each antenna element. The high diagnosis time required  by the techniques proposed in \cite{gen1}-\cite{music} and the limitations of the CS based techniques proposed in \cite{cs1}-\cite{cs5} motivate the development of new array diagnosis techniques suitable for mmWave systems.

In this paper, we develop low-complexity array diagnosis techniques for mmWave systems with large antenna arrays. These techniques account for practical assumptions on the mmWave hardware in which the analog phase shifters have constant modulus and quantized phases, and the number of RF chains is limited (assumed to be one in this paper).
The main contributions of the paper can be summarized as follows:
\begin{itemize}
\item We investigate the effects of random blockages on the far-field radiation pattern of linear uniform arrays. We consider both  partial and complete blockages. 
\item We derive closed-form expressions for the mean and variance of the far-field radiation pattern as a function of the antenna element blockage probability. These expressions provide an efficient means to evaluate the impact of  the number of antenna elements and  the antenna element blockage probability on the far-field radiation pattern.
\item We propose a new formulation for mmWave antenna diagnosis which relaxes the need for multi-location measurements, captures the sparse nature of  blockages, and enables efficient compressed sensing recovery.
\item We consider blockages at the transmit and/or receive antennas and propose two CS based array diagnosis techniques. These techniques identify the locations and the induced attenuation and phase shifts caused by unstructured blockages.
\item We exploit the two dimensional structure of mmWave antenna arrays and the correlation between the blocked antennas to further reduce the array diagnosis time when structured blockages exist at the receiver.
\item We evaluate the performance of the proposed array diagnosis techniques by simulations in a mmWave system setting, assuming that both the transmit and receive antennas are equipped with a single RF chain and 2-bit phase shifters.
\end{itemize}

The remainder of this paper is organized as follows. In Section \ref{sec:probform}, we formulate the array diagnosis problem and study the effects of random blockages on the far-field radiation pattern of linear arrays.  In Section \ref{sec:prop}, we introduce the proposed array diagnosis technique assuming a fault free transmit array and in Section \ref{sec:propj}, we introduce the proposed array diagnosis technique when faults are present at both the transmit and receive arrays. In Section \ref{sec:PA}  we provide some numerical results  and conclude our work in Section \ref{sec:con}.

\section{Problem Formulation} \label{sec:probform}

We consider a two-dimensional (2D) planar antenna array with $N_\text{x}$ equally spaced  elements along the x-axis and  $N_\text{y}$ equally spaced elements  along the y-axis; nonetheless, the model and the corresponding algorithms can be adapted to other antenna structures as well. Each antenna element is described by its position along the x and y axis, for example, the $(N_\text{x},N_\text{y})$th antenna refers to an antenna located at the $N_\text{x}$th position along the x-axis and the $N_\text{y}$th position along the y-axis.  The ideal far-field radiation pattern of this planar array in the direction $(\theta,\phi)$ is given by \cite{at} 
\begin{eqnarray}\label{cz1i}  
\hspace{-3mm } f(\theta,\phi) \hspace{-2.6mm }&=& \hspace{-3.7mm } \sum_{n=0}^{N_\text{y}-1} \hspace{-0.5mm } \sum_{m=0}^{N_\text{x}-1} \hspace{-1mm } w_{n,m}   e^{j   m \frac{2\pi d_x}{\lambda} \sin \theta \cos \phi}   e^{j n  \frac{2\pi d_y}{\lambda} \sin \theta \sin \phi}\hspace{-0.5mm },
\end{eqnarray}
where $d_x$ and $d_y$ are the antenna spacing along the x and y axis, $\lambda$ is the wavelength, and $w_{n,m}$ is the $(n,m)$th complex antenna weight.  

Let $\mathbf{a}_\text{x}(\theta,\phi) \in \mathcal{C}^{N_\text{x}\times 1}$ and $\mathbf{a}_\text{y}(\theta,\phi) \in \mathcal{C}^{N_\text{y}\times 1}$  be two vectors where the $m$th entry of $\mathbf{a}_\text{x}(\theta,\phi)$ is  $[\mathbf{a}_x(\theta,\phi)]_{m}= e^{j   m \frac{2\pi d_x}{\lambda} \sin \theta \cos \phi}$ and the $n$th entry of $\mathbf{a}_\text{y}(\theta,\phi)$ is  $[\mathbf{a}_y(\theta,\phi)]_{n} = e^{j   n \frac{2\pi d_y}{\lambda} \sin \theta \sin \phi} $.  Also, let the matrix $\mathbf{W}\in \mathcal{C}^{N_\text{y}\times N_\text{x}}$ be a matrix of antenna weights, where the $(n,m)$th entry of $\mathbf{W}$ is $[\mathbf{W}]_{n,m}=w_{n,m}$. Then (\ref{cz1i}) can be reduced to 
\begin{eqnarray}\label{cz2i} 
f(\theta,\phi)=  {\text{vec}{(\mathbf{W})}^\mathrm{T} }\mathbf{a}(\theta,\phi),
\end{eqnarray}
where vec$(\mathbf{W})$ is the $N_\text{x}N_\text{y} \times 1$ column vector obtained by stacking the columns of the matrix $\mathbf{W}$ on top of one another, the vector $\mathbf{a}(\theta,\phi)=\mathbf{a}_\text{x}(\theta,\phi) \otimes \mathbf{a}_\text{y}(\theta,\phi)$ is  the 1D array response vector, and the operator $\otimes$ represents the Kronecker product.   The formulation in (\ref{cz2i}) allows us to represent the 2D array as a 1D array, and as a result, simplify the problem formulation. 

\begin{figure}
		\begin{center}
\includegraphics[width=3.5in]{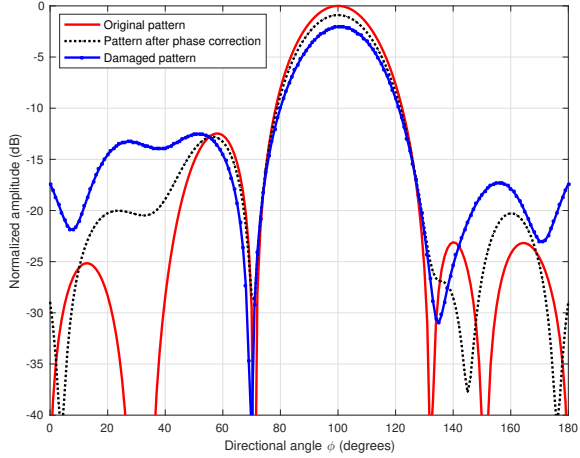}
\caption{Original and damaged beam patterns of a 16 element ($4 \times 4$) planar array with $\theta = 90$ degrees and $\frac{d_x}{\lambda}=\frac{d_y}{\lambda}=0.5$. The third, fifth and thirteenth array elements of the equivalent 1D  array (see (\ref{cz2i})) are blocked with $b_3=0.37+j0.22$, $b_5=-0.1+j0.34$, and $b_13=-0.64-j0.1$. Blockages result in an increase in the sidelobe level and a decrease in gain. Phase correction improves beam pattern, however, more complex precoder design is required for pattern correction.}
\label{fig:pat}
		\end{center}
	\end{figure}

In the presence blockages, the far-field radiation pattern of the array in (\ref{cz2i}) becomes
\begin{eqnarray}\label{bs1}
g(\theta,\phi)={\underbrace{\text{vec}{(\mathbf{W})}}_{\mathbf{x}}}^\mathrm{T}\underbrace{ (\mathbf{b} \circ  \mathbf{a}(\theta,\phi)) }_{\mathbf{z}},
\end{eqnarray}
where operator $\circ$  represents the Hadamard product, the vector $\mathbf{x} \in \mathcal{C}^{N_\text{x}N_\text{y}\times 1}$ is a vector of antenna weights and the vector $\mathbf{z} \in \mathcal{C}^{N_\text{x}N_\text{y}\times 1}$ is the equivalent array response vector. The $n$th entry of the vector  $\mathbf{b} \in \mathcal{C}^{N_\text{x}N_\text{y}\times 1}$ is defined by

  \begin{equation}\label{efbp1}
b_n  = \left\{
               \begin{array}{ll}
               \alpha_n, & \hbox{ if the $n$th element is blocked}  \\
               1, & \hbox{ otherwise,  } \\
               \end{array}
               \right.
\end{equation}
where $n = 1,..., N_\text{x}N_\text{y}$, $\alpha_n = \kappa_n e^{j\Phi_n}$,  $0 \leq \kappa_{n} \le 1$ and $0 \leq \Phi_{n} \leq 2\pi$ are the resulting absorption and scattering coefficients at the $n$th element.  A value of $\kappa_{n} = 0$ represents maximum absorption (or blockage) at the $n$th element, and the scattering coefficient $\Phi_{n}$ measures the phase-shift caused by the particle suspended on the $n$th element. This makes $b_n$ a random variable, i.e. $b_n=\alpha_n$ with probability $P_\text{b}$ if the $n$th antenna is blocked and $b_n=1$ with probability $1-P_\text{b}$ otherwise. It is clear from (\ref{bs1}) that blockages will change the array manifold and result in a distorted radiation pattern as shown in Fig. \ref{fig:pat}. The resulting pattern is a function of the number of the particles suspended on the array and their corresponding dielectric constants.

In Tables I and II, we summarize the effects of blockages on a linear array. Specifically, we tabulate the mean and variance of a distorted far-field radiation pattern of a linear array subject to blockages. The total number of blockages is assumed to be fixed, however, block locations and intensities (for the case of partial blockage) are assumed to be random.  For  ease of exposition, we study the effects on the azimuth direction only. A  similar  analysis  can  be  performed  for  the elevation pattern. Derivations of the results in Tables I and I can be found in \cite{mpv} and are omitted for space limitation. From Table I, we observe that both complete and partial blockages reduce the amplitude of the main lobe. This reduces the beamforming gain of the array. We also observe that complete blockages have no effect on the variance of the of the main lobe, and hence do not cause randomness in the main lobe. Random partial blockages, however, randomize the main lobe and lead to uncertainties in the mmWave channel. From Table II, we observe that complete and partial blockages distort the sidelobes of the far-field radiation pattern. Table II also shows that the variance of this distortion is a function of the blockage intensity and the antenna element blockage probability $P_\text{b}$.

\begin {table*}[t!]
\center
\caption{Mean and variance of the far-field beam pattern of a linear array steered at $\phi =\phi_\text{T}$ as a function of the antenna element blockage probability $P_\text{b}$ and the blockage coefficient $\alpha_n$. The blockage coefficient is constant if the array is subject to a single type of blockage, and random if the array is subject to multiple types of blockages.  Derivation of  results can be found in \cite{mpv} and are omitted for space limitations} 
\center
\begin{tabular}{ |p{5cm}||p{2.5cm}|p{1.5cm}|  }
 \hline
Fault Type &  Mean & Variance \\
 \hline
Complete blockage ($\alpha_{n\in \mathcal{I_\text{k}}}=0$) & $1-P_\text{b}$   & 0\\
Partial blockage  ($\alpha_{n\in \mathcal{I_\text{k}}}=\beta$)&  $1-P_\text{b}+P_\text{b}\beta$ &$0$ \\
Partial blockages ($\alpha_n$ random)&  $1-P_\text{b}+P_\text{b}\mathbb{E}[\alpha_n]$ &$P_\text{b} \text{var}[\alpha_n]$ \\
 \hline
\end{tabular}
\end{table*}

\begin {table*}[t!]
\caption{Mean and variance of the far-field beam pattern of a linear array steered at $\phi \not=\phi_\text{T}$ as a function of the antenna element blockage probability $P_\text{b}$ and the blockage coefficient $\alpha_n$; $\gamma = \frac{\pi d_x}{\lambda} (\cos (\phi_\text{}) - \cos (\phi_\text{T}) )$. Derivation of  results can be found in \cite{mpv} and are omitted for space limitation.} 
\center
\begin{tabular}{ |p{4.5cm}||p{5.2cm}|p{5.5cm}|  }
 \hline
Fault Type &  Mean & Variance \\
 \hline
Complete blockage ($\alpha_{n\in \mathcal{I_\text{k}}}=0$) & $(1-P_\text{b}) \frac{\sin ( N_x \gamma ) }{N_x \sin( \gamma)} e^{j (N_x-1) \gamma }$   & $ \frac{P_\text{b}}{N_x}\left(1-P_\text{b}\right)$\\
Partial blockage  ($\alpha_{n\in \mathcal{I_\text{k}}}=\beta$) &  $(1-P_\text{b}(1-\beta)) \frac{\sin ( N_x \gamma ) }{N_x \sin( \gamma)} e^{j (N_x-1) \gamma }$ &$  \frac{1}{N_x} (1-P_\text{b}+P_\text{b} |\beta|^2) -  \frac{1}{N_x} (1-P_\text{b}+P_\text{b}\beta)^2 $ \\
 Partial blockage ($\alpha_n$ random) &  $(1-P_\text{b}(1-\mathbb{E}[\alpha_n])) \frac{\sin ( N_x \gamma ) }{N_x \sin( \gamma)} e^{j (N_x-1) \gamma }$ &$ \frac{P_\text{b}}{N_x}(1- P_\text{b}+\mathbb{E} [\alpha^2_n] -P_\text{b}\mathbb{E} [\alpha_n]^2)$ \\
 \hline
\end{tabular}
\end{table*}

In the following sections, we propose several array diagnosis (or blockage detection/estimation) techniques for mmWave antenna arrays. Once array diagnosis is complete, the estimated attenuation and phase shifts caused by blockages can be used to calibrate the array.  Fig. 2, for example, shows the resulting pattern when phase correction is applied to the affected antenna elements. As shown, the resulting beam pattern is slightly improved, however, more complex precoder design is required to modify the excitation weights of the antenna elements and calibrate the array. The calibration process could for example focus on maximizing the beamforming gain and/or minimize the sidelobe level, and as a result, reduce the uncertainty in the mmWave channel. While the excitation weights of failed arrays can be easily modified in digital antenna architectures, additional hardware, e.g. RF chains, antenna switches, subarrays, etc., might be required to generate more degrees of freedom for the precoder design. Hybrid architectures (see e.g. \cite{h1} and \cite{h2}) can be used to modify the excitation weights of the failed array and also reduce the diagnosis time since each RF chain can now obtain independent measurements. Beam pattern correction and precoder design for failed arrays is beyond the scope of this work and is left for future work.  Before we proceed with the proposed techniques, we lay down the following assumptions: (i) The number of blockages is assumed to be small compared to the array size.  (ii) The channel between the transmitter and the receiver is line-of-sight (LoS) with a single dominant path. In the case of multi-path, the transmitter waits for a period $\tau$, where $\tau$  is proportional to the channel's delay spread, before it transmits the following training symbol. (iii) The transmit and receive array manifolds as well as the transmitter's angle-of-departure and the receiver's angle-of-arrival  are known at the receiver.  The AoA/D can be known a priori,  obtained by, for example, using prior sub-6 GHz channel information \cite{anum}, or provided by an infrastructure via a lower frequency control channel. (iv) Blockages remain constant for a time interval which is larger than the diagnosis time.

\section{Fault Detection at the Receiver}\label{sec:prop}
In the previous section, we showed that blockages distort the beam pattern of the array. To mitigate the effects of blockages, it is imperative to design reliable array diagnosis techniques that detect the fault locations and estimate the values of the blockage coefficients with minimum diagnosis time. Array diagnosis can be initiated after channel estimation. For example, the optional training subfield of the  SC PHY IEEE 802.11ad frame  could be utilized for periodic array diagnosis.  Since the system performance is greatly affected by the antenna architecture and precoder design, which we do not undertake in this work, we do not simulate the effect of CSI training loss in this paper and simply focus on the array diagnosis problem. Note, however, that beamforming at both the transmitter and receiver leads to larger coherence time \cite{ab1}, \cite{ab2} and the angular variation is typically an order of magnitude slower than the conventional coherence time \cite{ab1}. Since all AoDs/AoAs are assumed to be known a priori in this paper, the loss in CSI training time becomes unsubstantial.

In this section we propose three array diagnosis techniques. The first technique is generic in the sense that it does not exploit the block structure of blockages while the second and third techniques exploit the dependencies between blocked antenna elements to further reduce the array diagnosis time.


\subsection{Generic Fault Detection}\label{sec:GFD}
\begin{figure*}[t]
  \centering
  \begin{subfigure}[b]{0.5\linewidth}
    \centering\includegraphics[width=200pt]{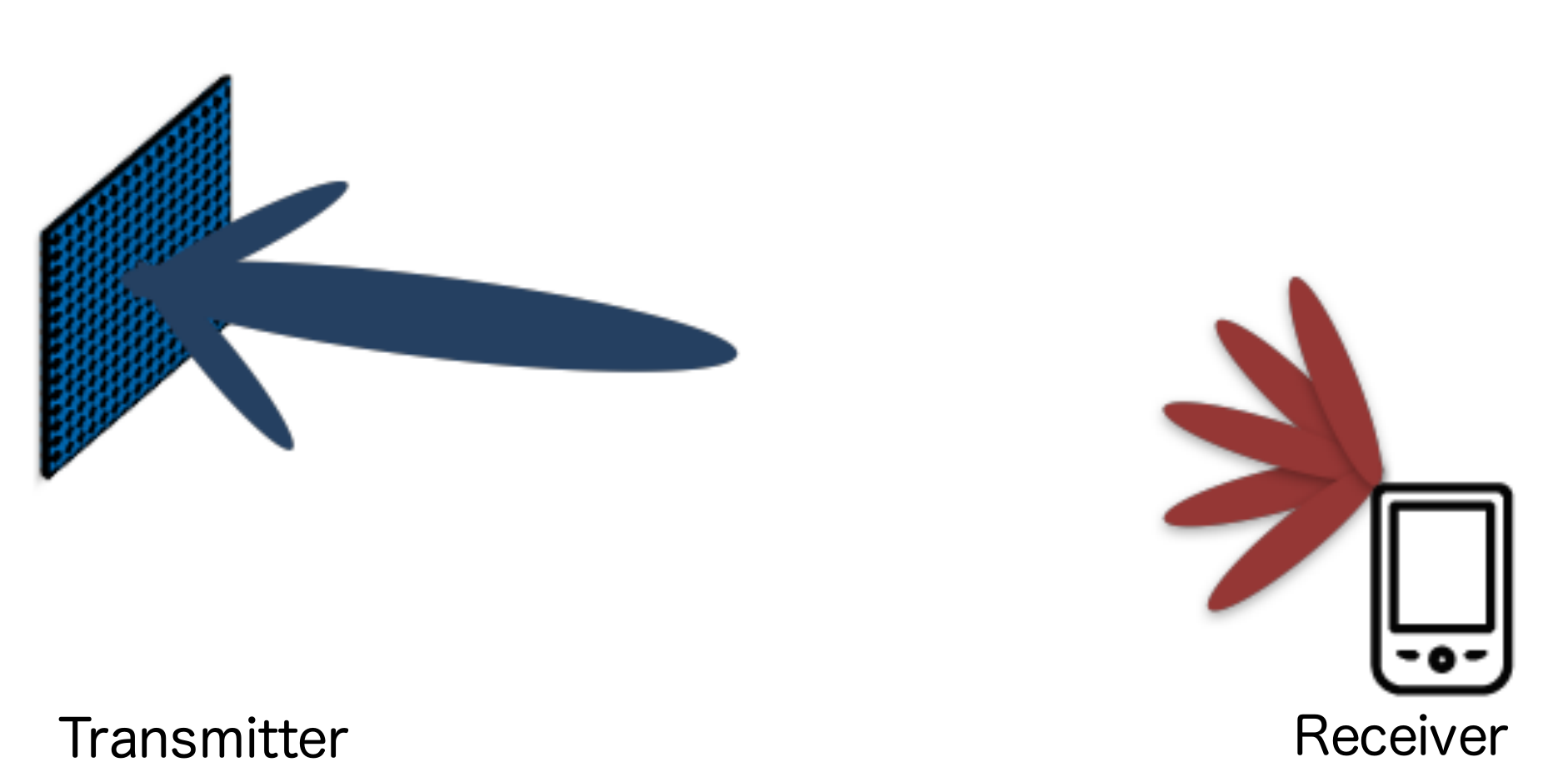}
    \caption{\label{fig:fig1}}
  \end{subfigure}%
  \begin{subfigure}[b]{0.5\linewidth}
    \centering\includegraphics[width=200pt]{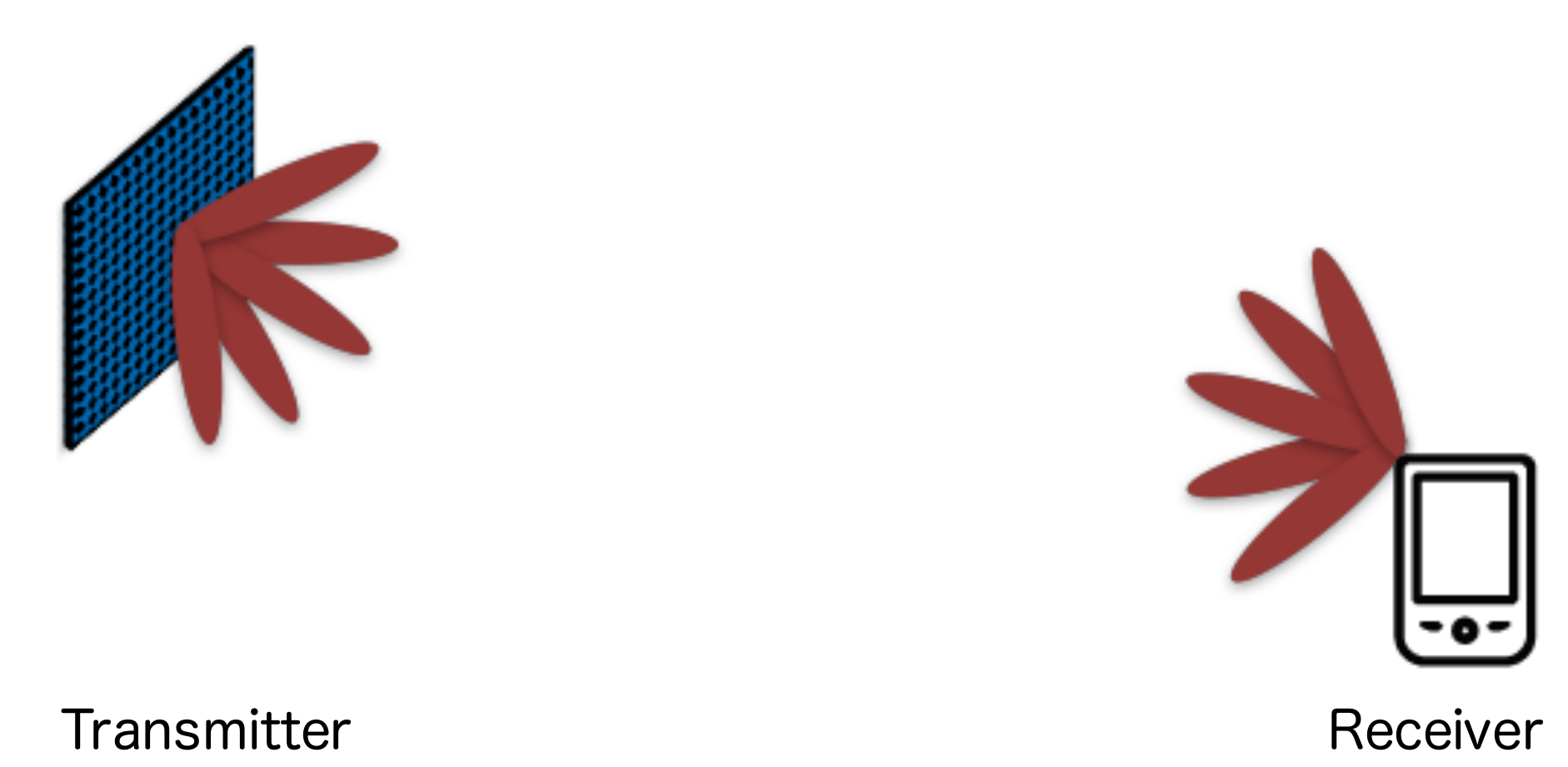}
    \caption{\label{fig:fig2}}
  \end{subfigure}
  \caption{An example of real time array diagnosis. (a) Diagnosis of the receiver array. Transmitter sends $K$ training symbols and the receiver receives each training symbol using a random receive beam. (b)  Joint diagnosis of both the transmit and the receive arrays. Transmitter sends training symbols using $k_\text{t}$ random transmit beams and the receiver receives each training symbol using $k_\text{r}$ random receive beams.} \label{adad}
\end{figure*}

To start the array diagnosis process,  the receiver with the AUT requests a transmitter, with known location, i.e. $\theta$ and $\phi$, to transmit $K$ training symbols (known to the receiver).  The receiver with the AUT generates a random beam to receive each training symbol as shown in Fig. \ref{adad}(a), i.e., random antenna weights are used at the AUT to combine each training symbol. Mathematically, the $k$th output of the AUT can be written as
\begin{eqnarray}\label{c4}
 {h}_k(\theta,\phi)=\sqrt{\rho} s \mathbf{x}_k^\mathrm{T}\mathbf{z}   + e_k,
\end{eqnarray}
where $k=1,...,K$, $\rho$ is the effective signal-to-noise ratio (SNR) which includes the path loss, $s=1$ is the training symbol, and $e_k \sim \mathcal{CN}(0,1)$ is the additive noise. The entries of the weighing vector $\mathbf{x}_k $ at the $k$th instant are chosen uniformly and independently at random. Equipped with the path-loss and angular location of the transmitter, the receiver generates the ideal pattern $f(\theta,\phi)$ using (\ref{cz1i}). Subtracting the ideal beam pattern from the received signal in (\ref{c4})  we obtain
\begin{align}
 \nonumber y_k &=  h_k(\theta,\phi) -f_k(\theta,\phi)=\mathbf{x}_k^\mathrm{T} (\mathbf{b} \circ  \mathbf{a}(\theta,\phi)) - \mathbf{x}_k^\mathrm{T} \mathbf{a}(\theta,\phi)  + \tilde{e}_k \\   \label{c2} &=\mathbf{x}_k^\mathrm{T}  (\mathbf{c} \circ  \mathbf{a}(\theta,\phi) ) + \tilde{e}_k,
 \end{align}
where $\tilde{e}_k = \frac{e_k}{\sqrt{\rho}}$, and the $n$th entry of the vector $\mathbf{c}$ is $[\mathbf{c}]_n=0$  when there is no blockage at the $n$th element, and $[\mathbf{c}]_n = b_{n} -1$ otherwise. Let $\mathbf{q} =  \mathbf{c} \circ  \mathbf{a}(\theta,\phi) $, after $K$ measurements we obtain

\begin{eqnarray}\nonumber
\hspace{-0mm}\underbrace{\left[ \begin{array}{c} y_1 \\ y_2\\ \vdots  \\y_K\end{array}
\right]}_{\mathbf{y}} \hspace{-1mm}= \hspace{-1mm}\underbrace{\left[
\begin{array}{cccc} x_{1,1} & x_{1,2} & \cdots & x_{1,N_\text{R}} \\
x_{2,1} & x_{2,2} & \cdots & x_{2, N_\text{R}} \\
\vdots & \vdots & \vdots & \vdots \\
x_{K,1} & x_{K,2} & \cdots & x_{K, N_\text{R}}
\end{array}
\right]}_{\mathbf{X}}
 \underbrace{\left[
\begin{array}{c} {q}_1 \\ {q}_2 \\ \vdots \\{q}_{ N_\text{R}} \end{array}
\right]}_{\mathbf{q}} \hspace{-1mm}+ \hspace{-1mm} \underbrace{\left[\begin{array}{c} \tilde{e}_1 \\ \tilde{e}_2 \\ \vdots \\\tilde{e}_K \end{array}\right]}_{\mathbf{e}}
\end{eqnarray}
or equivalently
\begin{eqnarray}\label{fb_modela}
\mathbf{y} = \mathbf{X}\mathbf{q} + \mathbf{e},
\end{eqnarray}
where $N_\text{R}=N_\text{x}N_\text{y}$ is the total number of antennas at the receiver. Assuming that the number of blocked antenna elements is small, i.e. $S\ll N_\text{R}$, the vector $\mathbf{q}$ in (\ref{fb_modela}) becomes sparse with $S$ non-zero elements that represent the locations of the blocked antennas. 

To  mitigate the effects of blockages, it is desired to first detect the locations of the blocked antennas and then the complex random variable $b_n$ with few measurements (or diagnostic time).  Note that the system in (\ref{fb_modela}) requires $K \ge N_\text{R}$ measurements to estimate the vector $\mathbf{q}$. While this might be acceptable for small antenna arrays, mmWave systems are usually equipped with large antenna arrays to provide sufficient link budget \cite{rap}.  Scaling the number of measurements with the number of antennas would require more measurements and would increase the array diagnosis time. In the following, we show how we can (i) detect  the locations of the blocked antennas, and (ii) estimate the complex blockage coefficients $b_n$ with $K \ll N_\text{R}$ measurements by exploiting the sparsity structure of the vector $\mathbf{q}$ under the assumption that blockages remain constant for a time interval which is larger than the diagnosis time. If the time interval is small, a hybrid architecture (see e.g. \cite{h1} and \cite{h2}) can be employed with multiple RF chains and each RF chain can obtain independent measurements. This reduces the diagnosis time by a factor of $N_\text{RF}$, where $N_\text{RF}$ is the number of RF chains.

\subsubsection{Sparsity Pattern Detection} \label{cs}
Compressive sensing theory permits efficient reconstruction of a sensed signal with only a few sensing measurements.  While there are many different methods used to solve sparse approximation problems (see, e.g., \cite{cP08}, \cite{spr}), we employ the least absolute shrinkage and selection operator (LASSO) as a recovery method. We adopt the LASSO since it does not require the support of the vector $\mathbf{q}$ to be known a priori. This makes it a suitable detection technique as blockages are random in general. The LASSO estimate of (\ref{fb_modela}) is given by \cite{cP08}
\begin{eqnarray}
\label{LASSO} \arg \min_{\boldsymbol{\nu} \in \mathbb{C}^{N_\text{R}\times 1}} \frac{1}{2}\| \mathbf{y} - \mathbf{X}\boldsymbol{{{\nu}}}\|_2^2 + \Omega \sigma \|\boldsymbol{{{\nu}}}\|_1,
\end{eqnarray}
where $\sigma$ is the standard derivation of the noise $\tilde{e}$, and $\Omega$ is a regularization parameter. The antennas weights, i.e. the entries of the matrix $\mathbf{X}$, are randomly and uniformly selected from the set $\{1+j, 1-j, -1+j, -1-j \}$ in this paper, i.e. weights can be applied to a mmWave antenna with 2-bit phase shifters. In the special case of 1-bit phase shifters, $\mathbf{X}$ becomes a Bernoulli matrix which is known to satisfy the coherence property with high probability \cite{cb1}, \cite{spr}, \cite{spr1}.
\subsubsection{Attenuation and Induced Phase Shift Estimation}\label{csls2}
Once the support $\mathcal{S}$, where $\mathcal{S}=\{n: q_n\neq0\}$,  one can apply estimation techniques such as least squares (LS) estimation to estimate and refine the complex coefficient $b_{n\in \mathcal{S}}$. To achieve this, the columns of $\mathbf{X}$ which are associated with the non-zero entries of $\mathbf{q}$ are removed to obtain $\mathbf{X}_\mathcal{S} \in \mathbb{C}^{K\times S}$. Hence, the vector $\mathbf{y}$ in equation (\ref{fb_modela}) can now be written as
\begin{eqnarray}\label{y2}
\mathbf{y}=\mathbf{X}_\mathcal{S}\mathbf{q}_\mathcal{S}+\mathbf{e},
\end{eqnarray}
where  $\mathbf{q}_\mathcal{S}$ is obtained by pruning the zero entries of $\mathbf{q}$. Since  $K>S$, the entries of  $\mathbf{q}_\mathcal{S}$  can be estimated via LS estimation. In particular, one can write the LS estimate after successful sparsity pattern recovery as \cite{LMMSE}
\begin{eqnarray} \label{24b}
\hat{\mathbf{q}}_{\mathcal{S}} = (\mathbf{X}_\mathcal{S}^*\mathbf{X}_\mathcal{S})^{-1} \mathbf{X}_\mathcal{S}^* \mathbf{y}
= \mathbf{q}_\mathcal{S} + \check{\mathbf{e}},
\end{eqnarray}
where $\hat{\mathbf{q}}_\mathcal{S}$ is a noisy estimate of $\mathbf{q}_\mathcal{S}$, and the entries of the output noise vector $\check{\mathbf{e}}$ are Gaussian random variables as linear operations preserve the Gaussian noise distribution. Note that the $n$th entry of the vector $\mathbf{q}$ is $[\mathbf{q}]_{n} = (b_n-1)a_n$, where $a_n$ is the $n$th entry of of the vector $\mathbf{a}(\theta,\phi)$ (see (\ref{c2})-(\ref{fb_modela})). Therefore, the estimated attenuation coefficient $\hat{\kappa}_{n\in\mathcal{S}} = |\frac{\hat{q}_{n\in\mathcal{S}}}{a_{n\in \mathcal{S}}}+1|$ and the estimated induced phase $\hat{\Phi}_{n\in \mathcal{S}} = \angle{\left(\frac{\hat{q}_{n\in\mathcal{S}}}{a_{n\in \mathcal{S}}}+1\right)}$.


\subsection{Exploiting the Block-Structure of Blockages} \label{sec:block1}
Due to the small antenna element size, it is likely that a suspended particle will block several neighboring antenna elements as shown in Fig. \ref{fig:ant}. This results in a block sparse structure which could be exploited to substantially reduce the number of measurements without scarifying robustness. In this section, we reformulate the CS problem to exploit this structure and reduce the array diagnosis time. To formulate the problem, we first rewrite $f(\theta,\phi)$  in (\ref{cz2i}) and $g(\theta,\phi)$ in (\ref{bs1})  as
\begin{eqnarray}\label{cbl1} 
f(\theta,\phi)=  {\text{vec}{(\mathbf{W})}^\mathrm{T} }{\text{vec}{(\mathbf{A})}},
\end{eqnarray}
and
\begin{eqnarray}\label{cbl1g} 
g(\theta,\phi)=  {\text{vec}{(\mathbf{W})}^\mathrm{T} }{\text{vec}{(\mathbf{A\circ B})}},
\end{eqnarray}
where $\mathbf{A}=\mathbf{a}_\text{y}(\theta,\phi) \mathbf{a}^\mathrm{T}_\text{x}(\theta,\phi)$ is the array response matrix, $\mathbf{W}$ is the weighting matrix,  and $\mathbf{B}$ is the sparse blockage matrix, i.e the entries of $\mathbf{B}$ are ``1'' in the case of no blockage and a random variable (see (\ref{efbp1})) in the case of a blockage. Substituting (\ref{cbl1}) and (\ref{cbl1g}) in (\ref{c4})-(\ref{c2}), the $k$th received measurement (after subtracting it form the ideal pattern) becomes
\begin{eqnarray}
\hspace{-10mm} {y}_{k} \hspace{-2mm}&=& \hspace{-2mm}{\text{vec}{(\mathbf{W}_k)}^\mathrm{T} }{\text{vec}{(\mathbf{A})}}- {\text{vec}{(\mathbf{W}_k)}^\mathrm{T} }{\text{vec}{(\mathbf{A\circ B})}} + \tilde{e}_k \\ \label{ykb10}  \hspace{-2mm}&=& \hspace{-2mm}  {\text{vec}{(\mathbf{W}_k)}^\mathrm{T} }{\text{vec}{(\mathbf{A}_\text{s})}}+\tilde{e}_k,
\end{eqnarray}
where $\mathbf{W}_k$ is the $k$th random weighting matrix, the innovation matrix $\mathbf{A}_\text{s}=\mathbf{A}-\mathbf{A}_\text{s}$ is sparse, and $e_k$ is the additive noise. Observe that the columns of the matrix $\mathbf{A}$ are either all 0's, in the case of no blockage, or contains a block of non-zero entries. We exploit this structure to reduce the number of measurements and as a result, reduce the array diagnosis time.

Let $\mathbf{X} = [{\text{vec}{(\mathbf{W}_1)}}, {\text{vec}{(\mathbf{W}_2)} }, \cdots, {\text{vec}{(\mathbf{W}_K)} }]^\mathrm{T}$ be the measurement matrix which consists of $K$ random antenna weights and $\mathbf{q} = {\text{vec}{(\mathbf{A}_\text{s})}}$, then after $K$ measurements the innovation vector  becomes
\begin{eqnarray}\label{yB}
\mathbf{y}_{\text{}} = \mathbf{X}_{\text{}}\mathbf{q}_{\text{}} + \mathbf{e}.
\end{eqnarray}
Observe that (\ref{yB}) is similar to (\ref{fb_modela}) with the exception that the new formulation allows the vector $\mathbf{q}$ to be block sparse. While the structure of the all 0 columns of $\mathbf{A}_\text{s}$ is known, the structure of the non-zero elements is unknown. This makes the block structure of the vector $\mathbf{q}$ in (\ref{yB}) random and a function of the number of blockages and their size. To complete the array diagnosis process, we employ the expanded block sparse Bayesian learning algorithm with bound optimization (EBSBL-BO) proposed in \cite{B0} to recover the block sparse matrix $\mathbf{q}$.  The EBSBL-BO algorithm exploits the intra-block correlation of the sparse vector to improve recovery performance without requiring prior knowledge of the block structure. From $\mathbf{q}$, the amplitude and induced phase shifts can be estimated as shown in Section \ref{csls2}.

\subsection{Extension to Complete Group-Blockages} \label{sec:block2}
\begin{figure}[t!]
		\begin{center}
\includegraphics[width=3in]{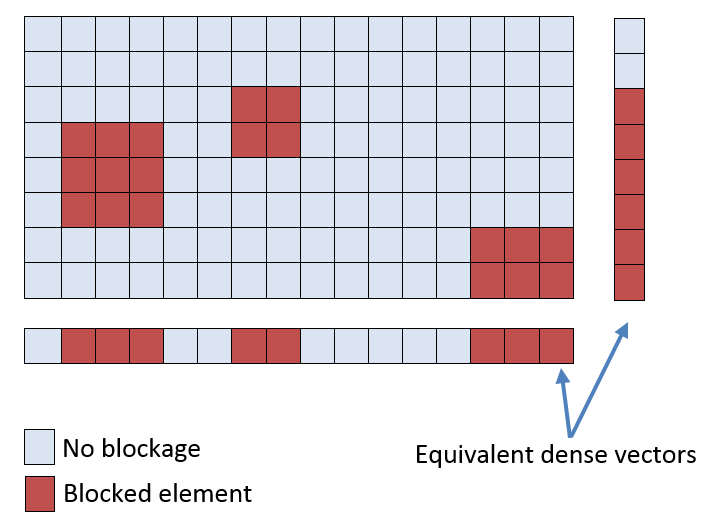}
\caption{An example of an innovation matrix (i.e. $\mathbf{A}_{\text{s}}=\mathbf{A}-\mathbf{A\circ B}$) of a 128 element antenna with 19 blocked/faulty elements. The equivalent dense vectors are obtained by summing the rows and columns of the matrix $\mathbf{A}_{\text{s}}$. Intersection of the indices of the non-zero elements (of the dense vectors)  correspond to the location of  potentially  blocked/faulty elements. }
			\label{fig:B1}
		\end{center}
	\end{figure}

When blockages are complete, i.e. $\alpha_n=0$, and span multiple neighboring antennas, the innovation matrix $\mathbf{A}_\text{s}$ in (\ref{ykb10}) becomes sparse with $J$ groups (or clusters) of non-zero entries at the locations of the faults as shown in Fig. \ref{fig:B1}. In this section, we exploit the structure of these faults and propose a technique that identifies the locations of these faults with just $N_\text{x}+N_\text{y}$ measurements provided that the number of groups is small, independent of the group size. Recall $N_\text{x}$ is the number of antennas along the x-axis and $N_\text{y}$ is the number of antennas along the y-axis. The idea is to decompose the matrix $\mathbf{A}_\text{s}$ into two dense vectors as shown in Fig. \ref{fig:B1}. The first vector is a weighted sum of the rows of $\mathbf{A}_\text{s}$ while the second vector is a weighted sum of the columns of $\mathbf{A}_\text{s}$. Since the number of unknown is now $N_\text{x}+N_\text{y}$, only $N_\text{x}+N_\text{y}$  measurements are required to recover the vectors. Once the vectors are recovered, the intersection of the non-zero elements of both vectors provides the location of potential faults. Equipped with the ideal matrix $\mathbf{A}$, the locations of the faults can be refined via an exhaustive search over all possible locations. In what follows, we formulate the problem and show how this is performed.

To formulate the problem, we rewrite ideal far-field radiation pattern in (\ref{cz1i}) as
\begin{eqnarray}\label{cz1irm}  
 f(\theta,\phi) = \mathbf{w}^{\mathrm{T}} \mathbf{A} \mathbf{p},
\end{eqnarray}
and the damaged pattern in (\ref{bs1}) as 
\begin{eqnarray}\label{cz1irmg}  
 g(\theta,\phi) = \mathbf{w}^{\mathrm{T}} (\mathbf{A}\circ \mathbf{B}) \mathbf{p},
\end{eqnarray}
where $\mathbf{w}\in \mathcal{C}^{N_\text{y}\times 1}$ and $\mathbf{p}\in \mathcal{C}^{N_\text{x}\times 1}$ represent the receive antenna weights, and the matrix $\mathbf{A}\in \mathcal{C}^{N_\text{y} \times N_\text{x}}$ is the antenna response matrix. Substituting (\ref{cz1irm}) and (\ref{cz1irmg}) in (\ref{c4})-(\ref{c2}), the $k$th received measurement (after subtracting it form the ideal pattern) becomes
\begin{eqnarray}\label{ykb10g}
\hspace{-8mm} {y}_{k} \hspace{-2mm}&=& \hspace{-2mm} {\mathbf{w}}^\mathrm{T}_k \mathbf{A}\mathbf{p}_k - {\mathbf{w}}^\mathrm{T}_k (\mathbf{A\circ B})\mathbf{p}_k + \tilde{e}_k = {\mathbf{w}}^\mathrm{T}_k \mathbf{A}_\text{s} \mathbf{p}_k+\tilde{e}_k,
\end{eqnarray}
To start the diagnosis process, the receiver obtains $N_\text{y}$ measurements using the weighting vector $\mathbf{w}$ while fixing the weighting vector $\mathbf{p}$. After $N_\text{y}$ measurements, the receiver obtains $N_\text{x}$ measurements using the weighting vector $\mathbf{p}$ while fixing the weighting vector $\mathbf{w}$. Mathematically, the innovation vector can be written as
\begin{eqnarray}\label{ygm}
\nonumber {y}_{1} &=& {\mathbf{w}^\mathrm{T}_1} \mathbf{A}_\text{s} \mathbf{p}_0+\tilde{e}_1\\
\nonumber \vdots && \hspace{10mm} \vdots \quad  \vdots\\
\nonumber {y}_{N_\text{y}} &=& \mathbf{w}^\mathrm{T}_{N_\text{y}} \mathbf{A}_\text{s} \mathbf{p}_0+\tilde{e}_{N_\text{y}}\\
\nonumber {y}_{N_\text{y}+1} &=& \mathbf{w}^\mathrm{T}_0 \mathbf{A}_\text{s} \mathbf{p}_{1}+\tilde{e}_{N_\text{y}+1}\\
\nonumber \vdots && \hspace{10mm} \vdots \quad  \vdots\\
\nonumber {y}_{N_\text{y}+N_\text{x}} &=&  \mathbf{w}^\mathrm{T}_0 \mathbf{A}_\text{s} \mathbf{p}_{N_\text{x}}+\tilde{e}_{N_\text{y}+N_\text{x}},
\end{eqnarray}
where the weighting vectors $\mathbf{w}_0 \in \mathcal{C}^{N_\text{y} \times 1}$ and $\mathbf{p}_0 \in \mathcal{C}^{N_\text{x} \times 1}$ consist of random weighting entries and are fixed throughout the diagnosis stage. Note that the term $\mathbf{A}_\text{s} \mathbf{p}_0$ represents a weighted sum of all the columns of matrix $\mathbf{A}_\text{s}$, and the term $\mathbf{w}^\mathrm{T}_0 \mathbf{A}_\text{s}$ represents a weighted sum of the rows of $\mathbf{A}_\text{s}$ (see Fig. (\ref{fig:B1})). To simplify above system of equations, let $\mathbf{x}_1=\mathbf{A}_\text{s} \mathbf{p}_0$, $\mathbf{x}_2=(\mathbf{w}^\mathrm{T}_0 \mathbf{A}_\text{s})^\mathrm{T}$, and the matrix $\boldsymbol{\Phi}$ be
\begin{eqnarray}
\boldsymbol{\Phi} = \left[
\begin{array}{cc} \mathbf{W} & \mathbf{0}_1 \\
\mathbf{0}_2 & \mathbf{P}  
\end{array}
\right],
\end{eqnarray}
where the weighting matrices $\mathbf{W} \in \mathcal{C}^{N_\text{y} \times N_\text{y}}$ and $\mathbf{P} \in \mathcal{C}^{N_\text{x} \times N_\text{x}}$  are both orthonormal matrices, the matrix $\mathbf{0}_1$ is an all zero matrix of size $N_\text{y} \times N_\text{x}$, the matrix $\mathbf{0}_2$ is an all zero matrix of size $N_\text{x} \times N_\text{y}$.
Then, the innovation vector can be simplified to 
\begin{eqnarray}\label{xxs}
\left[
\begin{array}{cc} {y}_1 \\
\vdots\\
{y}_{N_\text{y}+N_\text{x}}
\end{array}
\right] = \underbrace{\left[
\begin{array}{cc} \mathbf{W} & \mathbf{0}_1 \\
\mathbf{0}_2 & \mathbf{P}  
\end{array}
\right]}_{\boldsymbol{\Phi}}
\underbrace{\left[
\begin{array}{cc} \mathbf{x}_1 \\
\mathbf{x}_2
\end{array}
\right]}_{\mathbf{x}} + \left[
\begin{array}{cc} \tilde{e}_1 \\
\vdots\\
\tilde{e}_{N_\text{y}+N_\text{x}}
\end{array}
\right],
\end{eqnarray}
or equivalently
\begin{eqnarray}\label{1x}
\mathbf{y} = \boldsymbol{\Phi}\mathbf{x}+\mathbf{e}.
\end{eqnarray}
To recover the locations of the blockages, we estimate the vectors $\mathbf{x}_1$ and $\mathbf{x}_2$ in (\ref{xxs}) as follows
\begin{eqnarray}\label{1xe}
\hat{\mathbf{x}}= \boldsymbol{\Phi}^*\mathbf{y} = \boldsymbol{\Phi}^*\boldsymbol{\Phi}\mathbf{x}+\boldsymbol{\Phi}^*\mathbf{e}= \mathbf{x}+\acute{\mathbf{e}},
\end{eqnarray}
where $\hat{\mathbf{x}}$ is a noisy estimate of $\mathbf{x}$, the estimates $\hat{\mathbf{x}}_1 = [\hat{\mathbf{x}}]_{1:N_\text{y}} $ and $\hat{\mathbf{x}}_2=[\hat{\mathbf{x}}]_{(N_\text{y}+1):(N_\text{y}+N_\text{x})}$. The intersection of the indices of the non-zero elements in $\hat{\mathbf{x}}_1$ and $\hat{\mathbf{x}}_1$ correspond to potential blocked/fault locations. 

Using $\hat{\mathbf{x}}_1$ and $\hat{\mathbf{x}}_2$, we form the approximate binary matrix  $\tilde{\mathbf{B}}$ as follows
\begin{equation}\label{amow}
[\tilde{\mathbf{B}}]_{m,n}  = \left\{
               \begin{array}{ll}
               1, & \hbox{ if $[\mathbf{x}_1]_m$ and $[\mathbf{x}_2]_n$ are non-zero}  \\
               0, & \hbox{ otherwise.  } \\
               \end{array}
               \right.
\end{equation}
The binary matrix $\tilde{\mathbf{B}}$ can be refined as follows
\begin{eqnarray}\label{es}
\hat{\mathbf{B}}= \arg \min_{\mathbf{D} \in \{0,1\}} \bigg\| \hat{\mathbf{x}}-\left[
\begin{array}{cc} (\mathbf{A} - (\mathbf{A} \circ \mathbf{D})) \mathbf{p}_0 \\ 
(\mathbf{w}^{\mathrm{T}}_0  (\mathbf{A} - (\mathbf{A} \circ \mathbf{D})))^{\mathrm{T}}
\end{array}
\right]  \bigg\|_2,
\end{eqnarray}
where the matrix $\mathbf{D}$ is a submatrix  of the matrix $[\tilde{\mathbf{B}}]$. The zero entries of $\hat{\mathbf{B}}$ correspond to the locations of the blocked/faulty antennas.
\subsubsection*{Remarks}
\begin{enumerate}
\item For the special case of a single group blockage, i.e. $J=1$, the zero entries of the matrix $\tilde{\mathbf{B}}$ in (\ref{amow}) correspond to the locations of the faulty antennas and the search step in (\ref{es}) is not required.
\item When the number of groups increases, the computational complexity in (\ref{es}) increases and as a result, conventional compressed sensing recovery techniques might be more favorable in this case.
\item For large group sizes, the matrix $\mathbf{A}_s$ becomes dense and compressed sensing recovery techniques may fail in this case. The technique proposed in this section is able to identify the locations of the faulty antenna elements with just $N_\text{y}+N_\text{x}$ measurements. Even if compressed sensing recovery techniques succeed, the required number of measurements would be much higher than $N_\text{y}+N_\text{x}$.
\item When blockages are partial, the entries of the matrix $\mathbf{D}$ in (\ref{es}) would be complex instead of binary numbers, and the complexity of the exhaustive search step would be high if not prohibitive. A two-stage setting where the entries of $\tilde{\mathbf{B}}$ are independently examined might be favorable in this case.
\end{enumerate}

\section{Joint Fault Detection at the Transmitter and the Receiver}\label{sec:propj}

In the previous section, we assumed that the transmit antenna is free from blockages. When blockages exist at the transmit antenna, the receiver receives distorted training symbols and as a result, the technique proposed in the previous section will fail. In this section, we propose a detection technique that jointly detects blockages at both the transmitter and the receiver. For this technique, we assume that the receiver is equipped with the path-loss, angular location and array manifold of the transmitter.  To start the diagnosis process, the transmitter sends a training symbol $s=1$ using $k_\text{t}$ random antenna weights (assumed to be known by the receiver). The receiver generates $k_\text{r}$ random antenna weights to receives each training symbol, thus making the total number of measurements $K=k_\text{t}k_\text{r}$.  Let  $\mathbf{a}_{\text{T}}(\theta,\phi)=\mathbf{a}_\text{tx}(\theta,\phi) \otimes \mathbf{a}_\text{ty}(\theta,\phi)$ be  the $N_\text{T}\times 1$ 1D transmit array response vector. The $k$th received measurement at the receiver can be written as
\begin{eqnarray}\label{tx1}
y_k = \mathbf{w}_i^* (\mathbf{b\circ a}(\theta,\phi)) (\mathbf{b}_\text{T} \circ \mathbf{a}_{\text{T}}(\theta,\phi))^*\mathbf{f}_o+\tilde{e}_k,
\end{eqnarray}
where $\mathbf{w}_i$ is a random weighting vector at the receiver, $\mathbf{f}_o$ is a random weighting vector at the transmitter, and $\mathbf{b}_\text{T}$ is a vector of complex coefficients that result from the absorption and scattering caused by the particles blocking the transmit antenna array. After $K$ measurements the receiver obtains the following measurement matrix
\begin{eqnarray}\label{tx2}
\mathbf{Y}_\text{r}=  \mathbf{W}^*\mathbf{A}\mathbf{F}+\mathbf{E},
\end{eqnarray}
where $\mathbf{Y}_\text{r}$ is the received measurement matrix, $\mathbf{W} \in \mathbb{C}^{N_\text{R}\times k_\text{r}}$ is a random weighting matrix at the receiver, $\mathbf{A}  = (\mathbf{b\circ a}(\theta,\phi)) (\mathbf{b}_\text{T} \circ \mathbf{a}_{\text{T}}(\theta,\phi))^*$ is the $N_\text{R} \times N_\text{T}$ equivalent array response matrix, the matrix $\mathbf{F} \in \mathbb{C}^{N_\text{T}\times k_\text{t}}$ is a random weighting matrix at the transmitter, and $\mathbf{E}$ is the additive noise matrix. Equipped with the weighting matrices $\mathbf{F}$ and $\mathbf{W}$, the receiver generates the ideal  vector $\mathbf{Y}_\text{I}=\mathbf{W}^*\mathbf{A}_\text{I}\mathbf{F}$, where $\mathbf{A}_\text{I}=\mathbf{ a}(\theta,\phi)  \mathbf{a}^*_{\text{T}}(\theta,\phi)$, and subtracts it from $\mathbf{Y}_\text{r}$ in (\ref{tx2})  to obtain
\begin{eqnarray}\ \label{tx3a}
\mathbf{Y} = \mathbf{Y}_\text{r} - \mathbf{Y}_\text{I} 
=  \mathbf{W}^*\mathbf{A}_{\text{s}}\mathbf{F}+\mathbf{E} ,
\end{eqnarray}
where $\mathbf{A}_{\text{s}} \in \mathbb{C}^{N_\text{R}\times N_\text{T}}$ is a sparse matrix. The non-zero entries of the columns of $\mathbf{A}_{\text{s}}$ represent the indices of faulty antennas at the transmitter and the non-zero entries of the rows of $\mathbf{A}_{\text{s}}$ represent the entries of the faulty antennas at the receiver. To formulate the CS recovery problem, we vectorize the measurement matrix $\mathbf{Y}$ in (\ref{tx3a}) to obtain 
\begin{eqnarray}\label{tx5} \label{tx4}
\text{vec}(\mathbf{Y}) \hspace{-2mm}&=&\hspace{-2mm}\text{vec}( \mathbf{W}^*\mathbf{A}_{\text{s}}\mathbf{F})+\text{vec}(\mathbf{E})  \\
\hspace{-2mm}&=&\hspace{-2mm} \underbrace{(\mathbf{F}^{\mathrm{T}} \otimes\mathbf{W}^*)}_{\mathbf{U}} \underbrace{\text{vec}(\mathbf{A}_{\text{s}})}_{\mathbf{g}}+\underbrace{\text{vec}(\mathbf{E})}_{\mathbf{e}},
\end{eqnarray}
where the vector $\mathbf{y}=\text{vec}(\mathbf{Y})$, the matrix $\mathbf{U}\in \mathbb{C}^{K\times N_\text{T}N_\text{R}}$ is the effective CS sensing matrix, and the vector $\mathbf{g}\in \mathbb{C}^{N_\text{T}N_\text{R}\times 1}$ is the effective sparse vector.   Note  that if the matrices $\mathbf{W}$ and $\mathbf{F}$ in (\ref{tx5}) both satisfy the coherence property, examples include Gaussian, Bernoulli and Fourier matrices \cite{cb1}, then the matrix $\mathbf{U}=\mathbf{W}\otimes \mathbf{F}$ also satisfies the coherence property and it can be applied in standard CS techniques.  For simplicity, the entries of $\mathbf{W}$ and $\mathbf{F}$  are chosen uniformly and independently at random from the set $\{1+j, 1-j,  -1+j, -1-j \}$ in this paper. This corresponds to 2-bit phase shifters at both the transmitter and the receiver.

\subsection{Sparsity Pattern Detection and Least Squares Estimation} \label{csls}
 The LASSO estimate of (\ref{tx5}) is given by 
 \begin{eqnarray}
\label{LASSO2} \arg \min_{\boldsymbol{\nu} \in \mathbb{C}^{N_\text{T}N_\text{R}\times 1}} \frac{1}{2}\|  {\mathbf{y}} -  {\mathbf{U}}\boldsymbol{{{\nu}}}\|_{2}^2 + \Omega \sigma_\text{e} \|\boldsymbol{{{\nu}}}\|_{1},
\end{eqnarray}
where  $\sigma_\text{e}$ is the standard derivation of the noise $e$. Once the support $\mathcal{S}$, where $\mathcal{S}=\{i:  {g}_i\neq0\}$, of the vector $ {\mathbf{g}}$ is estimated, the columns of $ {\mathbf{U}}$ which are associated with the non-zero entries of $ {\mathbf{g}}$ are removed to obtain $ {\mathbf{U}}_\mathcal{S}$. Hence, the vector $ {\mathbf{y}}$ in (\ref{tx5}) becomes
\begin{eqnarray}\label{tx6}
 {\mathbf{y}}= {\mathbf{U}}_\mathcal{S} {\mathbf{g}}_\mathcal{S}+ {\mathbf{e}},
\end{eqnarray}
where  $ {\mathbf{g}}_\mathcal{S}$ is obtained by pruning the zero entries of $ {\mathbf{g}}$. The LS estimate after successful sparsity pattern recovery becomes
\begin{eqnarray}\label{tx6b} 
\hat{\mathbf{g}}_{\mathcal{S}} = ( {\mathbf{U}}_\mathcal{S}^* {\mathbf{U}}_\mathcal{S})^{-1}  {\mathbf{U}}_\mathcal{S}^*  {\mathbf{y}}
=  {\mathbf{g}}_\mathcal{S} +  {\mathbf{e}},
\end{eqnarray}
where $\hat{\mathbf{g}}_\mathcal{S}$ is a noisy estimate of $ {\mathbf{g}}_\mathcal{S}$.

\subsection{Attenuation and Induced Phase Shift Estimation}
Let $\mathbf{r}$ be a vector of  size $N_\text{T}N_\text{R}\times 1$ and its $i$th entry is $[\mathbf{r}]_i = [\mathbf{g}_\mathcal{S}]_i \text{ in } (\ref{tx6b})$, if $i\in\mathcal{S}$ and $r_{i}=0$ otherwise. Reshaping $\mathbf{r}$ into $N_\text{R}$ rows and $N_\text{T}$ columns we obtain an estimate $\hat{\mathbf{A}}_s$ of the sparse matrix ${\mathbf{A}}_s$ in (\ref{tx3a}). The non-zero columns of $\hat{\mathbf{A}}_s$ represent the IDs of the transmit array faulty antennas, and the non-zero rows  of $\hat{\mathbf{A}}_s$ represent the IDs of the receive array faulty antennas. Let the set $\mathcal{I}_\text{r}$ contain the indices of the zero rows of $\hat{\mathbf{A}}_s$ and the  set $\mathcal{I}_\text{t}$ contain the indices of the zero columns of $\hat{\mathbf{A}}_s$. Removing the rows associated with the IDs of the faulty receive antennas from $\hat{\mathbf{A}}_s$ we obtain the matrix $\mathbf{A}_r =  [\hat{\mathbf{A}}_s]_{\mathcal{I}_\text{r},:}$,  and the sparse vector that represents the faulty transmit antennas becomes
\begin{eqnarray}\label{txtxo}
\hat{\mathbf{q}}_\text{t} =\frac{(\mathbf{a}_r(\theta,\phi)^*{\mathbf{A}}_r)^*}{\|\mathbf{a}_r(\theta,\phi)\|_2^2},
\end{eqnarray}
where the vector $\mathbf{a}_r(\theta,\phi)$ results from selecting the $\mathcal{I}_\text{r}$ entries from the vector $ \mathbf{a}(\theta,\phi)$ in (\ref{tx1}). Based on (\ref{txtxo}), the estimated attenuation coefficient and induced phase  of the $i$th  antenna element at the transmit array becomes $\hat{\kappa}_{\text{t},i} = |\frac{[\hat{\mathbf{q}}_\text{t}]_i}{[\mathbf{a}_\text{T}(\theta,\phi)]_i}+1|$ and  $\hat{\Phi}_{\text{t},i} = \angle{\left(\frac{[\hat{\mathbf{q}}_\text{t}]_{i}}{[\mathbf{a}_\text{T}(\theta,\phi)]_i}+1\right)}$.

Similarly, removing the columns associated with the IDs of the faulty antennas from $\hat{\mathbf{A}}_s$ we obtain the matrix ${\mathbf{A}}_t = [\hat{\mathbf{A}}_s]_{:,\mathcal{I}_\text{t}}$, and the sparse vector that represents the faulty receive antennas becomes
\begin{eqnarray}\label{txrxx}
\hat{\mathbf{q}}_\text{r} =\frac{{\mathbf{A}}_t \mathbf{a}_t(\theta,\phi)}{\|\mathbf{a}_t(\theta,\phi)\|_2^2},
\end{eqnarray}
where the vector $\mathbf{a}_t(\theta,\phi)$ results from selecting the $\mathcal{I}_\text{t}$ entries from the vector $ \mathbf{a}_\text{T}(\theta,\phi)$ in (\ref{tx1}).
From (\ref{txrxx}),  the estimated attenuation coefficient and induced phase  of the $i$th  antenna element at the receive array becomes   $\hat{\kappa}_{\text{r},i} = |\frac{[\hat{\mathbf{q}}_\text{r}]_i}{[\mathbf{a}_\text{}(\theta,\phi)]_i}+1|$ and $\hat{\Phi}_{\text{r},i} = \angle{\left(\frac{[\hat{\mathbf{q}}_\text{r}]_{i}}{[\mathbf{a}_\text{}(\theta,\phi)]_i}+1\right)}$.

\section{Numerical Validation } \label{sec:PA}

In this section, we conduct numerical simulations to evaluate the performance of the proposed techniques.  We consider a 2D planar array, with $\frac{d_x}{\lambda}=\frac{d_y}{\lambda}=0.5$, that experiences  random and independent blockages with probability $P_\text{b}$. To generate the random blockages, the values of $\kappa$ and $\Phi$ in (\ref{efbp1}) are chosen uniformly and independently at random from the set  $\{i \in \mathcal{R}: 0 \leq i \leq 1\}$ and $\{0,..,2\pi\}$ respectively.  We adopt the success probability, i.e. the probability that all faulty antennas are detected, and the  normalized mean square error (NMSE) as a performance measure to quantify the error in detecting the blocked antenna locations and estimating the corresponding blockage coefficients  ($\kappa$ and $\Phi$). The NMSE is defined by
\begin{eqnarray}\label{cc1}
\text{NMSE} = \frac{  \| \mathbf{v}-\hat{\mathbf{v}} \|^2_2 }{   \| \mathbf{v} \|^2_2  }.
\end{eqnarray}

When blockages only exist at the receiver, $\mathbf{v} =  \mathbf{c} \circ  \mathbf{a}(\theta,\phi)$ (see (\ref{c2})), and the $i$th entry of the estimated vector $\hat{\mathbf{v}}$  is $[\hat{\mathbf{v}}]_{i\in \mathcal{S}}  = [\hat{\mathbf{q}}]_{i\in \mathcal{S}}$ in (\ref{24b}) and zero otherwise.  When blockages exist at both the receiver and the transmitter,  ${\mathbf{v}} =    (\mathbf{b}_\text{T}\circ \mathbf{a}_\text{T})-\mathbf{a}_\text{T}$ (see (\ref{tx1}) and (\ref{tx3a})) and $\hat{\mathbf{v}} =  \hat{\mathbf{q}}_t$ in (\ref{txtxo}) when detecting blockages at the transmitter array, and  ${\mathbf{v}} =    (\mathbf{b}_\text{}\circ \mathbf{a}_\text{})-\mathbf{a}_\text{}$ and $\hat{\mathbf{v}} =  \hat{\mathbf{q}_r}$ in (\ref{txrxx}) when detecting blockages at the receiver array.  To implement the LASSO, we use the function {\it{SolveLasso}} included in the {\it{SparseLab}} toolbox \cite{SL}.  As a benchmark, we compare the NMSE of the proposed techniques with the NMSE of the Genie aided LS estimate which indicates the optimal estimation performance when the exact locations of  the faulty antennas is known, i.e., the support $\mathcal{S}$ in (\ref{24b}) and (\ref{tx6b}) is assumed to be provided by a Genie. 

%
%
%

\begin{figure*}
\centering
\begin{subfigure}{.5\textwidth}
  \centering
\includegraphics[width=3.5in]{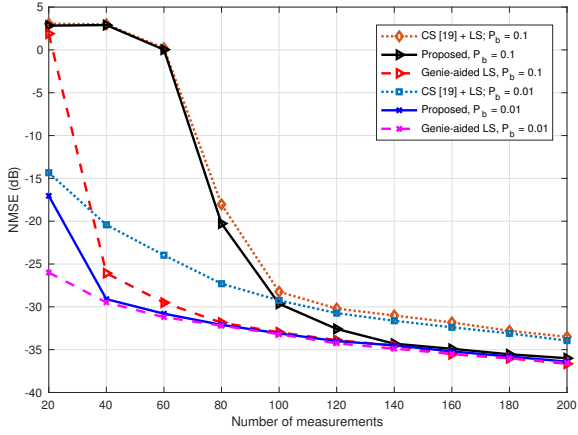}
\caption{$N_\text{R}= 16 \times 16 =256$.}
\label{fig:nmse1}
\end{subfigure}%
\begin{subfigure}{.5\textwidth}
  \centering
  \vspace{-1mm}
\includegraphics[width=3.4in]{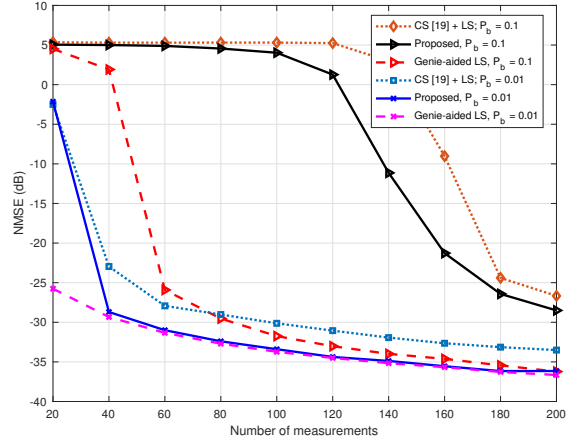} {\vspace{-2mm}
\caption{$N_\text{R}= 16 \times 32 = 512$.}
\label{fig:nmse2}}
\end{subfigure}
\caption{Detection and estimation of faults in a 2D receive planar array subject to random partial blockages with different blockage probability $P_\text{b}$; $\rho = 10$ dB, and  $N_\text{R}= 16 \times 16 =256$. Blockages do not occur in groups and the technique proposed in Section \ref{sec:GFD} is used for diagnosis.}
\label{fig:nmseb}
\end{figure*}

\begin{figure}[t!]
		\begin{center}
\includegraphics[width=3.7in]{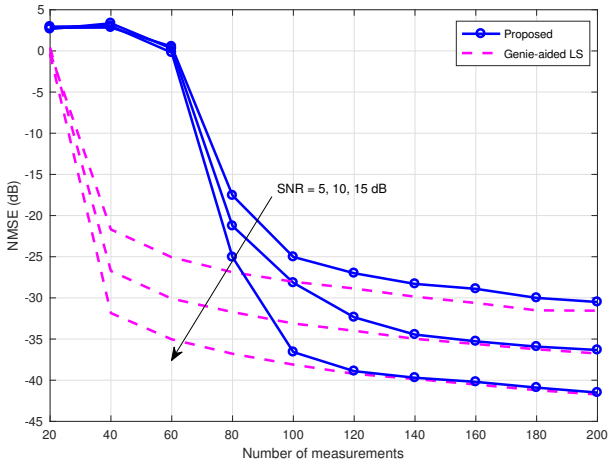}
\caption{ Detection and estimation of faults in a 2D receive planar array subjected to random partial blockages with different receiver SNR; $P_\text{b} = 0.1$, and $N_\text{R}= 16 \times 16 = 256$.  Blockages do not occur in groups and the technique proposed in Section \ref{sec:GFD} is used for diagnosis.}
			\label{fig:nmse3}
		\end{center}
	\end{figure}

\begin{figure}[t!]
		\begin{center}
\includegraphics[width=3.7in]{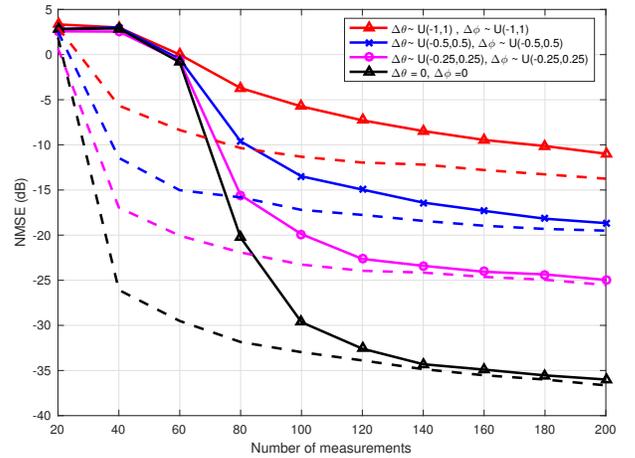}
\caption{Detection and estimation of faults in a 2D receive planar array subject to random partial blockages with AoD estimation errors; $P_\text{b}=0.1$; $\rho = 10$ dB, and  $N_\text{R}= 16 \times 16 =256$. Blockages do not occur in groups and the technique proposed in Section \ref{sec:GFD} is used for diagnosis. Dashed lines represent Gene aided LS estimation.}
			\label{fig:nmseAS}
		\end{center}
	\end{figure}

\begin{figure}[t!]
		\begin{center}
\includegraphics[width=3.7in]{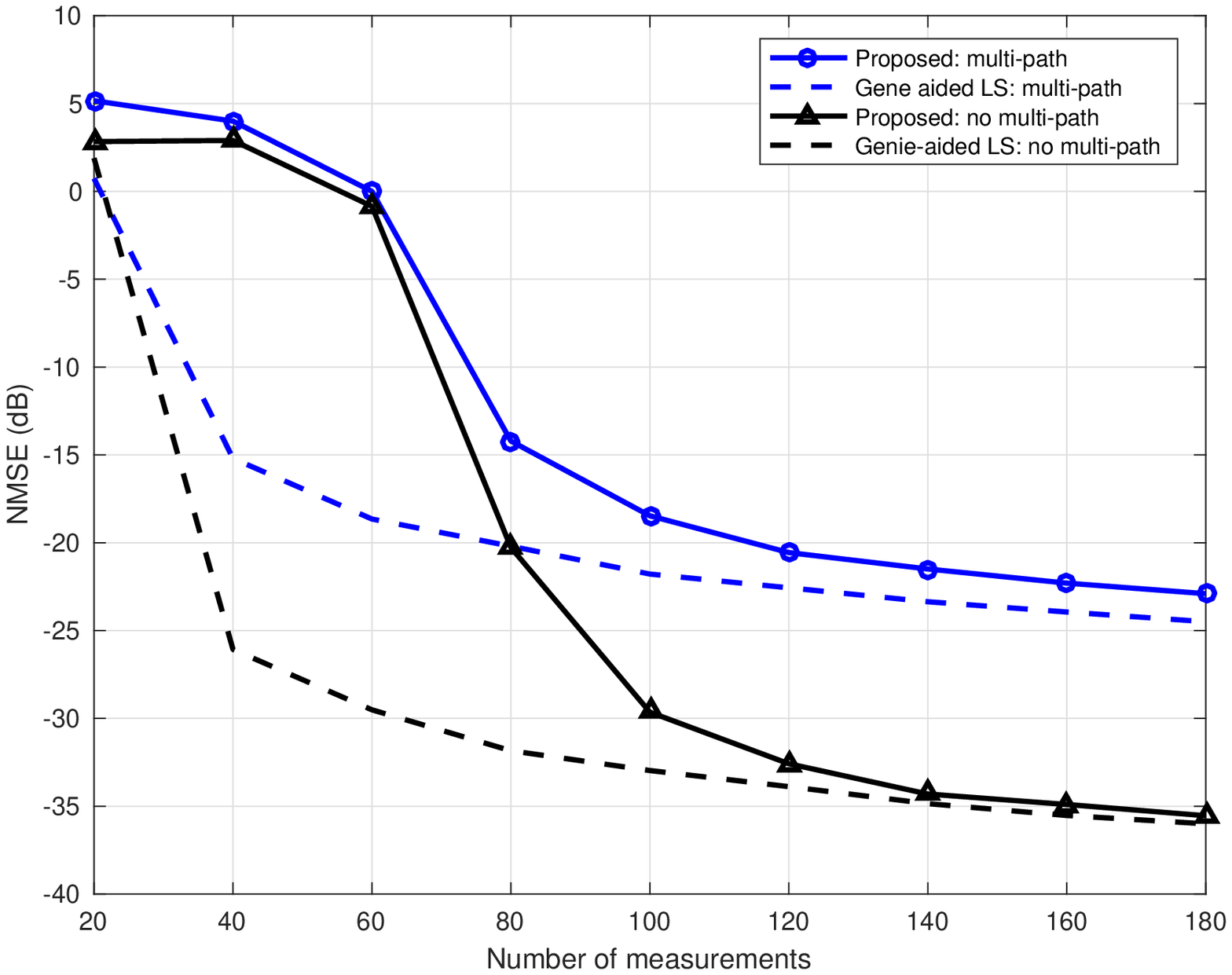}
\caption{Detection and estimation of faults in a 2D receive planar array subject to random partial blockages in the presence of random indirect (or scattered) path; $P_\text{b}=0.1$; $\rho = 10$ dB,  $N_\text{R}= 16 \times 16 =256$. Total number of path = 3 (one direct path plus two random path with random delays uniformly distributed across the diagnosis period),  and 90\% of the energy is located in the direct path.}
			\label{fig:nmseMP}
		\end{center}
	\end{figure}

In Figs. \ref{fig:nmseb}-\ref{fig:nmseMP} we consider faults at the receiver antenna array and assume that the transmitter antenna array is fault free. To study the effect of the number of measurements (or diagnosis time) on the performance of the proposed fault detection technique in Section \ref{sec:GFD}, we plot the NMSE when the receive array is subject to blockages with different probabilities in  Fig. \ref{fig:nmse1}.  For all cases, we observe that the NMSE decreases with increasing number of measurements $K$. The figure also shows that for sufficient number of measurements (on the order of $K\sim \mathcal{O}(P_\text{b}N_\text{R} \log N_\text{R})$), the NMSE of the proposed technique matches the NMSE obtained by the Genie-aided LS technique. This indicates for sufficient number of  measurements, the proposed technique successfully detects the locations of the blocked antennas and the corresponding blockage coefficients with $K \ll N_\text{R}$ measurements.  The figure also shows that as the blockage probability increases, more measurements are required to reduce the NMSE. The reason for this is that as the blockage probability increases, the average number of blocked antennas increases as well. Therefore, more measurements are required to estimate the locations of the blocked antennas and the corresponding blockage coefficients. In the event of large number of blockages or fast varying blockages, a hybrid antenna architecture with a few RF chains can be adopted to reduce the array diagnosis time.

  For comparison, we plot  the performance of the CS-based diagnosis technique proposed in \cite{cs1} which requires measurements to be randomly taken at $N_\text{R}$ locations, in Figs. \ref{fig:nmse1} and {\ref{fig:nmse2}}. This technique is chosen as it is based on analog beamforming, and therefore, it can be applied to mmWave systems.  For both $P_\text{b}=0.01$ and $P_\text{b}=0.1$, the proposed technique provides a lower NMSE while requiring lower number of measurements. For instance, to obtain a target NMSE of -30dB with $P_\text{b}=0.01$, the proposed technique requires 45 measurements while the algorithm proposed in \cite{cs1} requires 110 measurements, taken at 110 independent locations. This makes the proposed algorithm superior as it is able to obtain lower NMSE with lower number of measurements without the need for taking measurements at multiple locations.

To examine the effect of the array size on the number of measurements, we plot the NMSE of the proposed technique with  $N_\text{R}= 512$ in  Fig. {\ref{fig:nmse2}}. The figure shows that the NMSE decreases with increasing number of measurements $K$. Nonetheless, this decrease occurs at a lower rate when compared to the case when $N_\text{R}= 256$ in Fig. \ref{fig:nmse1}. This is particularly observed for higher blockage probabilities.   The reason for this is that as the array size increases, the average number of blocked antennas increases as well. Therefore, more measurements are required to estimate the locations of the blocked antennas and the corresponding blockage coefficients. Similar to the case when $N_\text{R}= 256$, the proposed technique results in lower NMSE with lower diagnosis time when compared to \cite{cs1}. This is mainly due to the CS sensing matrix which in this paper is optimized to satisfy the coherence property, thereby requiring lower number of measurements.

The effect of the SNR on the performance of the proposed technique is shown in  Fig. \ref{fig:nmse3}.  The figure shows that for sufficient number of measurements, the NMSE obtained by the proposed technique approaches the NMSE obtained by the Genie-aided techniques. The figure also shows that required number of measurements is a function of the receive SNR. For instance, for a receive SNR of 15 dB and 120 measurements, the NMSE of the proposed technique is similar to the NMSE obtained by the Genie-aided technique. As the SNR decreases to 5 dB, more than 200 measurements are required to match the NMSE of the Genie aided technique. To reduce the number of measurements, one can increase the receive SNR by either placing more antennas at the transmitter to increase the array gain or reduce the transmitter-receiver distance to minimize the path-loss. 
 
To investigate the effect of imperfect AoD/AoA estimation on the NMSE performance, we plot the NMSE of the proposed technique in Fig. \ref{fig:nmseAS} under the assumption of imperfect  azimuth and elevation angles of departure, i.e. $\phi = \phi+\Delta \phi$, and $\theta = \theta+\Delta \theta$, where $\Delta\phi$ and $\Delta \theta$ are uniformly distributed random variables.  Fig. \ref{fig:nmseAS}  shows that angular deviations (as small as $\pm 0.25^{\circ}$) result in a 10 dB loss in NMSE performance and this loss increases with larger angular deviations. The reason for this NMSE increase is due to the AoD/AoA mismatch which increases the system noise.  Fig. \ref{fig:nmseAS}  also shows that for sufficient number of measurements,  the NMSE of the proposed technique matches the NMSE of the Gene-aided LS estimation technique (which assumes perfect knowledge of the location of blockages). This suggests that majority of the NMSE results from blockage coefficient estimation errors rather than blockage location detection errors.

In Fig. \ref{fig:nmseMP} we study the impact of random multipath on the NMSE performance. Similar to the imperfect AoD/AoA case, multipath introduces interference which increases the noise floor of the system and, as a result, deteriorate the NMSE performance. We consider  a direct path and 2 multipath with random delays and gains. For this setup, 90\% of the received signal energy is contained in the direct path and the scattered path delays are uniformly distributed across the diagnosis time. As expected, Fig. \ref{fig:nmseMP} shows that the NMSE is lower in the presence of multipath and the NMSE does not decrease with increasing number of measurements. This is mainly attributed to multipath interference which is treated as noise in this paper.

	  	\begin{figure}[t!]
		\begin{center}
\includegraphics[width=3.8in]{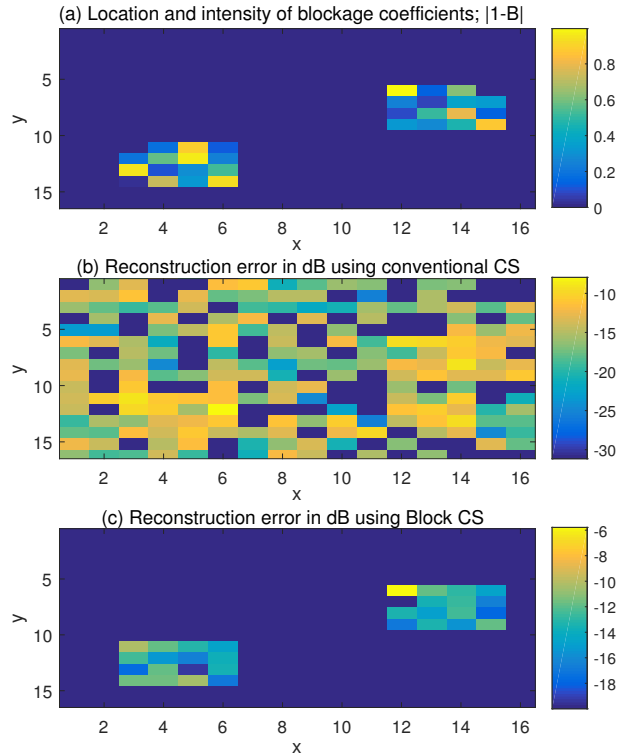}
\caption{ Comparison of the proposed array diagnosis techniques in detecting antenna faults in an  $N_\text{R}= 16 \times 16 = 256$ element antenna; (a) Location and intensity of blockages incident on the array ($|\mathbf{1}-\mathbf{B}|$, where $\mathbf{1}$ is an all-ones matrix). (b) Reconstruction error ($|\mathbf{B}-\hat{\mathbf{B}}|$) in dB when ignoring the block structure. The reconstructed matrix is denoted by $\hat{\mathbf{B}}$. (c) Reconstruction error when exploiting the block structure of blockages using EBSBL-BO+LS. For (b) and (c), $\rho$ = 10 dB, number of blocks is 2, block size $J$ = 16, and the number of measurements are fixed to 130.}
			\label{fig:B2}
		\end{center}
	\end{figure}

	  	\begin{figure}[t!]
		\begin{center}
\includegraphics[width=3.7in]{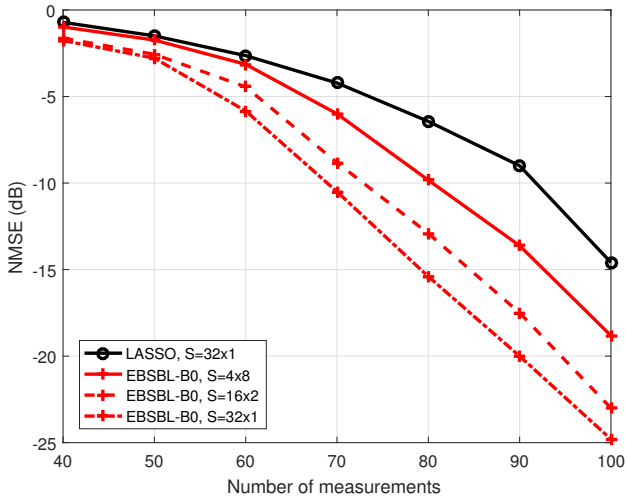}
\caption{ Comparison of the proposed techniques in detecting faults in a 2D receive planar array subjected to random group blockages. The proposed technique that ignores the block structure uses conventional LASSO as a recovery method, while the technique that exploits the block structure uses the EBSBL-B0 \cite{B0} algorithm for recovery. Blockages affect a group of $S=\Gamma \times J$ antenna elements, where $\Gamma$ is the group size and $J$ is the number of groups. Each antenna element within a group experiences a random blockage intensity; $\rho = 10$ dB, $N_\text{R}= 16 \times 16 = 256$.}
			\label{fig:B12}
		\end{center}
	\end{figure}

	  	\begin{figure}[t!]
		\begin{center}
\includegraphics[width=3.7in]{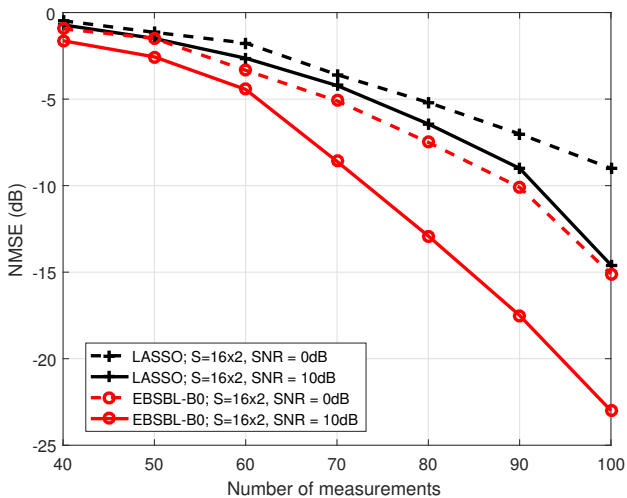}
\caption{ Comparison of the proposed techniques in detecting faults in a 2D receive planar array subjected to random group blockages and different noise levels. Each antenna element within a group experiences a random blockage intensity; $N_\text{R}= 16 \times 16 = 256$.}
			\label{fig:B13}
		\end{center}
	\end{figure}

%
%
%
%
%
%
%

\begin{figure*}
\centering
\begin{subfigure}{.5\textwidth}
  \centering
\includegraphics[width=3.5in]{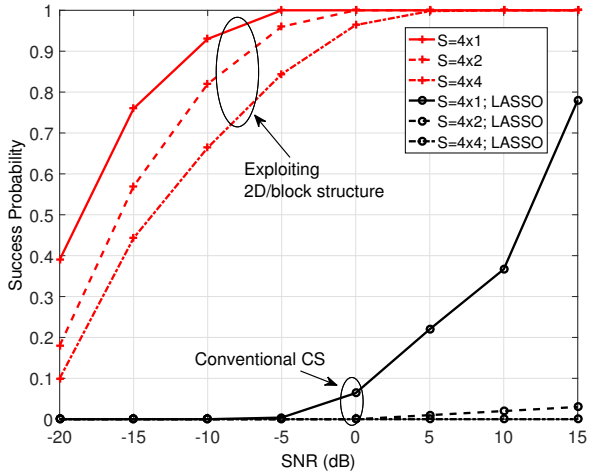}
\caption{Fixed block-size.}
\label{fig:B14}
\end{subfigure}%
\begin{subfigure}{.5\textwidth}
  \centering
\includegraphics[width=3.5in]{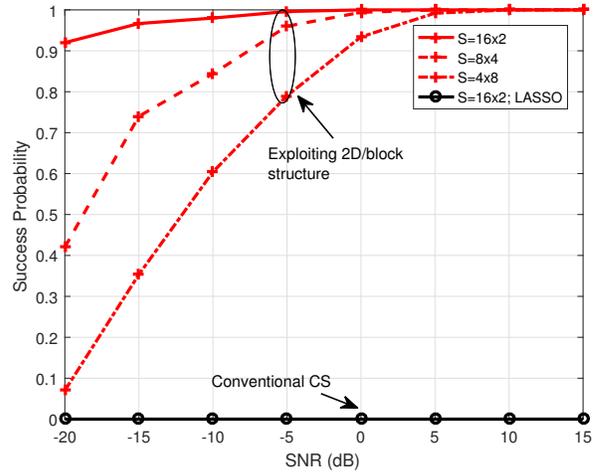}
\caption{Variable block size.}
			\label{fig:B15}
\end{subfigure}
\caption{Detection of faults in a 2D receive planar array subjected to complete (non-partial) blockages that occur in groups as a function of the receive SNR; $K = 16$ measurements, and $N_\text{R}= 8 \times 8 = 64$.}
			\label{fig:B152}
\end{figure*}

   	\begin{figure}
		\begin{center}
\includegraphics[width=3.7in]{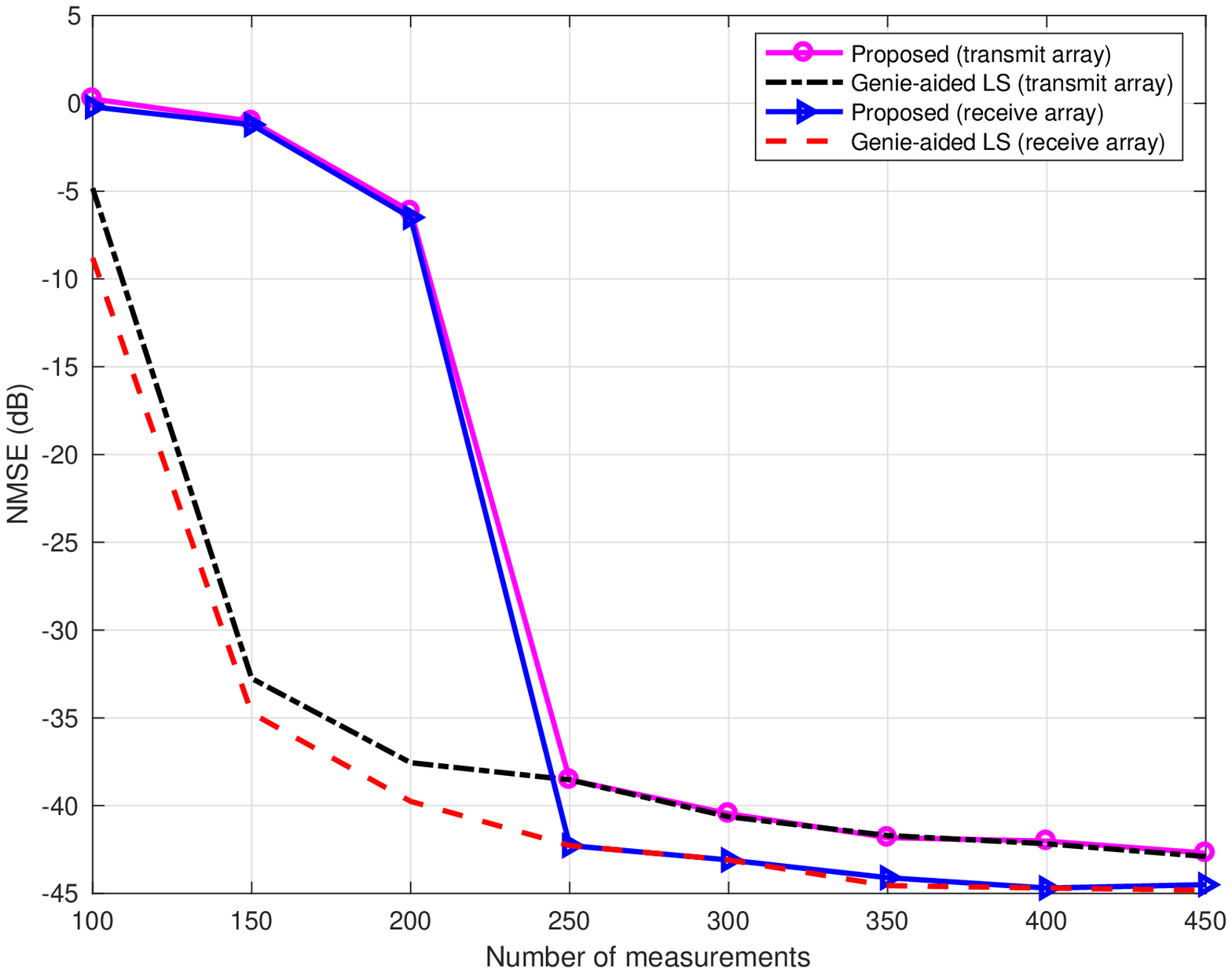}
\caption{Joint detection of faults in 2D transmit and receive planar arrays both subject to random partial blockages; $P_\text{b} = 0.1$ for both arrays, $\rho = 0$ dB,   $N_\text{R} = 16$,  $N_\text{T} = 32,$ and  $N_\text{T}N_\text{R} = 512$.}
			\label{fig:nmse6}
		\end{center}
	\end{figure}

In Figs. \ref{fig:B2}-\ref{fig:B152} we consider blockages that span a group of neighboring antennas. More specifically, in Fig. \ref{fig:B2} we consider blockages that affect 16 neighboring antennas. To model the random shapes of the blocking particles, we assume that each antenna element within a block experiences a random blockage intensity, i.e. the values of $\kappa$ and $\Phi$ in (\ref{efbp1}) are chosen uniformly and independently at random from the set  $\{i \in \mathcal{R}: 0 \leq i \leq 1\}$ and $\{0,..,2\pi\}$. In Fig. \ref{fig:B2}(a), we plot an example of a 256 element array subject to 2 blockages and each blockage affects a group of 16 antennas with random intensities. In Fig. \ref{fig:B2}(b), we plot the reconstruction error when the proposed technique is used to detect and estimate the blockage coefficients without exploiting the block structure of blockages. As shown, the proposed technique, using conventional CS, identifies locations of the blocked antennas, however, it also results in some false positives which can be reduced by increasing the number of measurements.  In Fig. \ref{fig:B2}(c), we plot the reconstruction error when we exploit the block structure of blockages to detected and estimate the blockage coefficients. As shown, this technique leverages the dependencies between the values and locations of the blockages in its recovery process and as a result it minimizes false positives and reduces the number of required measurements.

To highlight the benefit of exploiting correlation between the blocked antenna elements, we compare the NMSE achieved when (i) ignoring the block structure of the blockages and using conventional CS recovery, and (ii) exploiting the block structure of blockages by implementing the technique proposed in Section \ref{sec:block1} in Fig. \ref{fig:B12} and Fig. \ref{fig:B13}. Specifically, we consider random blockages and each blockage spans $\Gamma$ antennas with random blockage intensity. For array diagnosis, we plot the performance of the proposed techniques with and without exploiting the block structure. For $\Gamma=32$ and $J=1$, Fig. \ref{fig:B12} shows that lower NMSE can be achieved when exploiting the block structure. For fixed number of blockages, Fig. \ref{fig:B12} shows that the performance gap decreases with decreasing block size $\Gamma$. As the block size decreases, the correlation between the blocked antenna elements diminishes and hence the performance of this technique comes closer to that achieved by techniques that ignore the block structure of blockages. In Fig. \ref{fig:B13} we compare the performance of the proposed techniques when exploiting and ignoring the block structure of blockages for different receive SNRs. For all cases, the plots show a clear performance benefit when exploiting the block structure in the array diagnosis process.

In Figs. \ref{fig:B14} and \ref{fig:B15}, we consider complete blockages and compare the performance of the technique proposed in Section \ref{sec:block2} with conventional CS recovery techniques (LASSO) for fixed number of measurements $K=16$.  Fig. \ref{fig:B14} shows that the proposed technique achieves higher success probability when compared to conventional CS recovery techniques. The success probability of the proposed technique is highest when the number of blocks is $J=1$. When the number of blocks increases, however, the search space increases (see (\ref{es})), and as a result, we observe a performance hit especially at low SNR. In Fig. \ref{fig:B15} we fix the total number of blockages and study the impact of the block size on the success probability. Fig. \ref{fig:B15} shows that for fixed number of blockages, higher success probability is achieved for larger block sizes and the success probability decreases with increasing number of groups. As the number of groups increases, the search space in (\ref{es}) increases, and in the presence of noise, false alarms could occur. This reduces the success probability. Note that in Figs. \ref{fig:B14} and \ref{fig:B15}, conventional CS techniques fails since the required number of required measurements is $K>S \log N_\text{y}N_\text{x}$ which is much more than $N_\text{y}+N_\text{x}$.

    	\begin{figure}[t]
		\begin{center}
\includegraphics[width=3.7in]{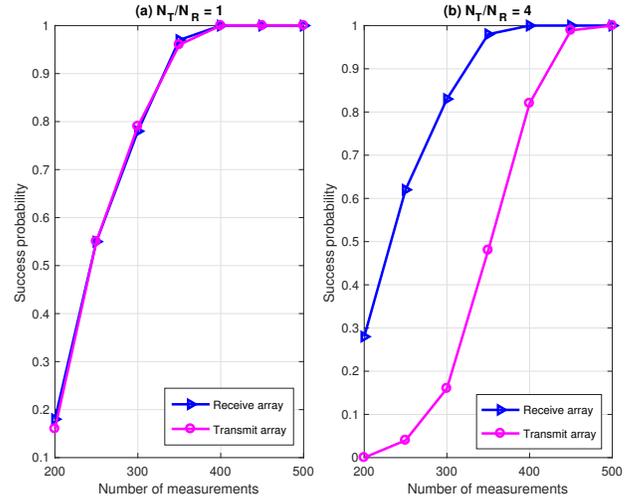}
\caption{Joint detection of faults in  2D transmit and receive planar arrays all subject to constant blockages; $P_\text{b} = 0.1$ for both arrays, $[\mathbf{b}_\text{T}]_i \in \{0,1\}$, $[\mathbf{b}]_i, \in \{0,1\}$, and  $N_\text{T}N_\text{R} = 1024$. In (a) $N_\text{R} = 32$,  and $N_\text{T} = 32,$ and in (b) $N_\text{R} = 16$,  and $N_\text{T} = 64$.}
			\label{fig:pontnr}
		\end{center}
	\end{figure}

In Figs. \ref{fig:nmse6}-\ref{fig:pontnr}, we assume that both the transmit and receive arrays are subject to random blockages. To analyze the effect of the number of measurements (or diagnosis time) on the NMSE performance when jointly detecting faults on both the transmit and receive arrays, we plot the NMSE of the technique proposed in Section \ref{sec:propj} and the Genie-aided technique in Fig. \ref{fig:nmse6}.  For both cases, we observe that the NMSE decreases with increasing number of measurements $K$, and for sufficient $K$, the NMSE of the proposed technique matches the NMSE obtained by the Genie-aided technique. The figure also shows that the NMSE of the receive array is lower than the NMSE of the transmit array. This is due to the transmit-receive array size difference. For a fixed blockage probability $P_\text{b}$, larger array sizes encounter more blockages on average, and therefore, require more measurements to detect faults.

The success probability when detecting faulty antennas in both the transmit and the receive arrays is plotted in Fig. \ref{fig:pontnr}. Fig. \ref{fig:pontnr} (a) shows that for both the transmit and receive array, the success probability increases with increasing number of measurements. Fig. \ref{fig:pontnr} (b) shows that lower number of measurements are required to detect faulty antennas at the receiver when the transmit array size is larger than the receive array size. As the transmit array size increases, the average number of blocked antennas increases as well, thereby requiring more  measurements to successfully detect the fault locations.

\section{Conclusions} \label{sec:con}
In this paper, we investigated the effects of blockages on mmWave linear antenna arrays. We showed that both complete and partial blockages distort the far-field beam pattern of a linear array and partial blockages result in a higher beam pattern variance when compared to complete blockages. To detect blockages, we proposed several compressed sensing based array diagnosis techniques. The proposed techniques do not require the AUT to be physically removed and do not require any hardware modification. When faults exist at the receiver only, we showed that the proposed techniques reliably detect the locations of the blocked antennas, if any, and estimate the corresponding attenuation and phase-shift coefficients caused by the blocking particles. Moreover, we showed that the dependencies between the blocked antennas can be exploited to further reduce estimation errors and lower the diagnosis time.  When faults exist at both the receiver and the transmiter, we showed that reliable detection and estimatation of blockages can be achievded. Nonetheless, high number of measurements are required in this case, even if compressed sensing is used. For all cases, the estimated coefficients can be used to calculate new antenna excitations weights to recalibrate the transmit/receive antennas. Due to their reliability and low diagnosis time, the proposed techniques can be used to perform real-time mmWave antenna array diagnosis at the receiver and enhance the mmWave communication link.

{}
\end{document}